\RequirePackage{fix-cm}
\documentclass[smallextended]{svjour3}       % onecolumn (second format)
\smartqed  % flush right qed marks, e.g. at end of proof
\usepackage{graphicx}
\usepackage{natbib}
\usepackage{amsmath}
\usepackage{physics}
\usepackage{amssymb}
\usepackage{xcolor}
\usepackage{enumerate}
\usepackage[]{algorithm2e}

\usepackage{physics,soul}
\newcommand*{\vertbar}{\rule[-1ex]{0.5pt}{2.5ex}}
\newcommand{\Rey}{\mathit{Re}}
\newcommand{\Nfft}{N_{\mathrm{FFT}}}
\newcommand{\Nblk}{N_{\mathrm{blk}}}

\newcommand{\Novlp}{N_{\mathrm{ovlp}}}
\newcommand{\tol}{\mathit{tol}}
\newcommand{\lb}{\mathit{lb}}
\newcommand{\ub}{\mathit{ub}}
\newcommand{\ii}{\mathrm{i}}
\newcommand{\ee}{\mathrm{e}}
\newcommand{\eqn}[1]{(\ref{#1})}
\DeclareMathSymbol{\shortminus}{\mathbin}{AMSa}{"39}
\DeclareMathOperator*{\argmax}{arg\,max} 
\newcommand{\ldb}{\mathopen{\lbrack\!\lbrack}} 
\newcommand{\rdb}{\mathclose{\rbrack\!\rbrack}}

\begin{document}

%\title{Bispectral mode decomposition of nonlinear flows %\thanks{Grants or other notes
%%about the article that should go on the front page should be
%%placed here. General acknowledgments should be placed at the end of the article.}
%}
%\subtitle{Discovery of triadic interactions from data}

\title{Bispectral mode decomposition of nonlinear flows}


\author{Oliver T. Schmidt}        % \and
%        Second Author %etc.
%}


\institute{O. T. Schmidt \at
              Department of Mechanical and Aerospace Engineering\\ University of California San Diego, La Jolla, CA, USA \\
              Tel.: +1 (858) 246-5818\\
              \email{oschmidt@ucsd.edu}           %  \\
           %\and
           %S. Author \at
           %   second address
}

\date{Received: date / Accepted: date}

\maketitle

\begin{abstract}
Triadic interactions are the fundamental mechanism of energy transfer in fluid flows. This work introduces bispectral mode decomposition as a direct means of educing flow structures that are associated with triadic interactions from experimental or numerical data. Triadic interactions are characterized by quadratic phase coupling which can be detected by the bispectrum. The proposed method maximizes an integral measure of this third-order statistic to compute modes associated with frequency triads, as well as a mode bispectrum that identifies resonant three-wave interactions. Unlike the classical bispectrum, the decomposition establishes a causal relationship between the three frequency components of a triad. This permits the distinction of sum- and difference-interactions, and the computation of interaction maps that indicate regions of nonlinear coupling. Three examples highlight different aspects of the method. Cascading triads and their regions of interaction are educed from direct numerical simulation data of laminar cylinder flow. It is further demonstrated that linear instability mechanisms that attain an appreciable amplitude are revealed indirectly by their difference-self-interactions. Applicability to turbulent flows and noise-rejection are demonstrated on particle image velocimetry data of a massively separated wake. The generation of sub- and ultraharmonics in large eddy simulation data of a transitional jet is explained by extending the method to cross-bispectral information.

%\nocite{schmidt2020bispectral}
\end{abstract}

\section{Introduction}
\label{intro}

Triadic interactions result from the quadratic nonlinearity of the Navier-Stokes equations. They are the fundamental mechanism of energy transfer in fluid flows and manifest, in Fourier space, as triplets of three  wavenumber vectors, $\{\vb{k}_j,\vb{k}_k,\vb{k}_l\}$, or frequencies, $\{f_j,f_k,f_l\}$, that sum to zero:
\begin{subequations}
\begin{eqnarray}
\vb{k}_j\pm\vb{k}_k\pm\vb{k}_l&=&\vb{0},\label{eqn:triad_wavenum}\\ f_j\pm f_k\pm f_l&=&0.\label{eqn:triad}
\end{eqnarray}
\end{subequations}
For clarity, we denote by $\{\cdot\}$ multiplets of frequency or wavenumber, and by $(\cdot)$ index multiplets. 

\begin{figure}
  \includegraphics[width=0.45\textwidth,trim=0 0cm 0cm 0cm,clip]{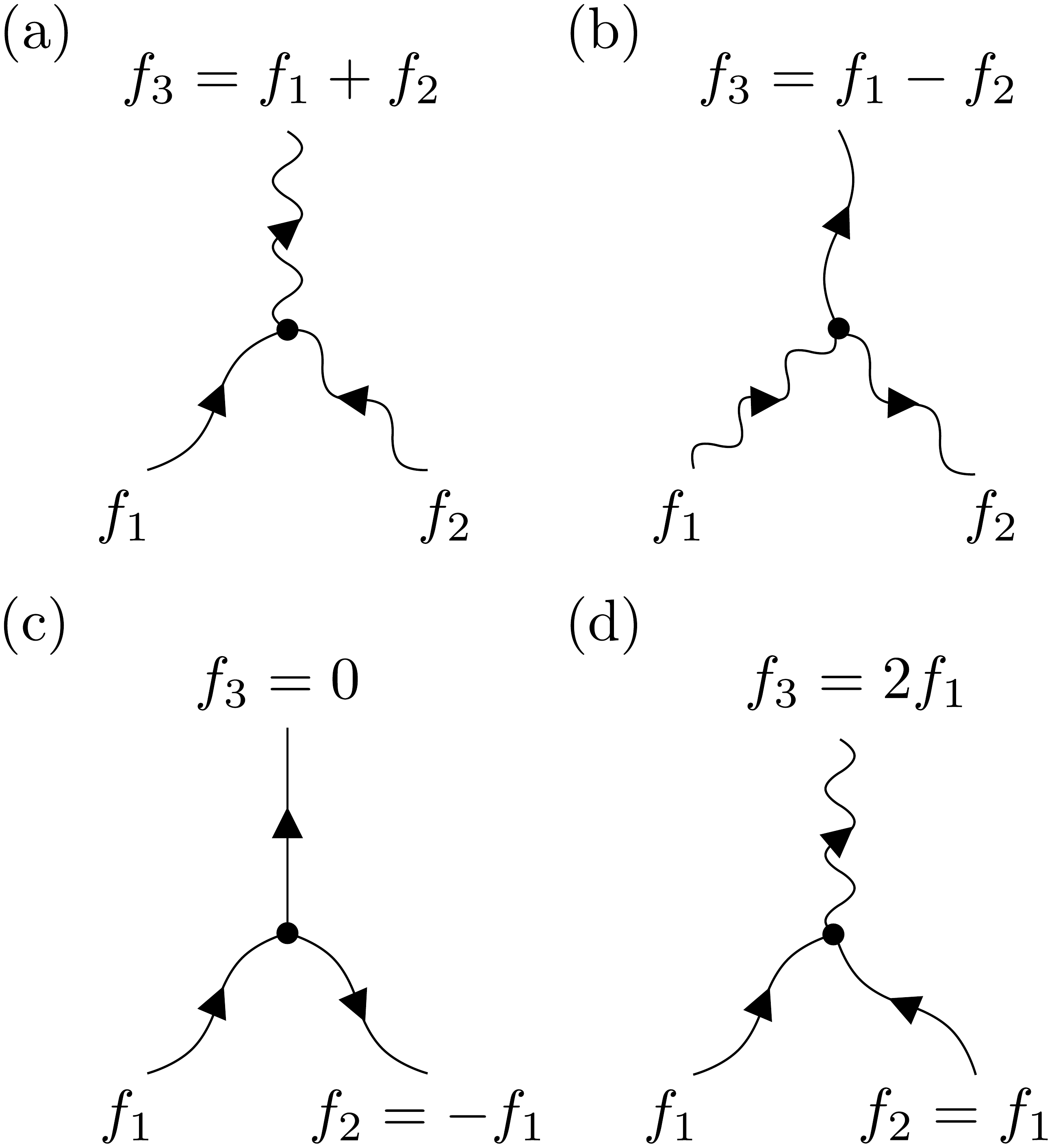}
\caption{Illustration of typical frequency triads: (a) generic sum-interaction; (b) generic difference-interaction; (c) mean-flow deformation generated by difference-self-interaction; (d) harmonic generated by sum-self-interaction. Directions of $f_1$ and $f_2$ indicate  sum ($-\!\!\!\!\blacktriangleright\!\!\!\!\!-\!\!\bullet\!\!\!-\!\!\!\!\!\blacktriangleleft\!\!\!\!-$) and difference ($-\!\!\!\!\blacktriangleright\!\!\!\!-\!\!\!\bullet\!\!\!-\!\!\!\!\blacktriangleright\!\!\!\!-$) interactions.}
\label{fig:triadic_interaction_Feynman}       % Give a unique label
\end{figure}
The zero-sum condition implies that triads form triangles in wavenumber- and frequency-space. A way to conceptually visualize these three-wave interactions is presented in figure \ref{fig:triadic_interaction_Feynman}. Since the early work of \cite{phillips1960dynamics} on weak resonant interactions of gravity waves on the surface of deep water, interaction theory has vastly improved our understanding of nonlinear wave phenomena. The turbulent cascade, which describes the transfer of energy from large to small scales of motion, is probably the most prominent consequence of triadic interactions. 

Following the seminal work by \cite{kraichnan1967inertial,kraichnan1971inertial}, the role of triad interactions and triad truncation in homogenous turbulence has been studied by numerous authors including \cite{waleffe1992nature} and \cite{moffatt2014note}. Using direct numerical simulation data, \cite{domaradzki1990local} investigated the energy transfer between scales through triad interaction in homogeneous and isotropic turbulence. Later, \cite{domaradzki1992nonlocal} proposed a self-similar relation that predicts the $k^{-5/3}$ scaling of the energy spectrum in the inertial range and the $k^{-2}e^{-ak}$ scaling in the far dissipation range. By encoding the condition for triadic resonance into a combination matrix, \cite{cheung2014exact} were able to recover the energy scaling of the inertial range directly from the Navier-Stokes equations. Triads play a similarly important role in the laminar-turbulent transition process. This was demonstrated by \cite{craik1971non}, who showed that resonant triads of Tollmien-Schlichting waves provide an efficient mechanism for rapid transition in wall-bounded shear flows. Recently, \citet{rigas2020non} demonstrated that this transition process can be modeled by a limited number of harmonics and their triadic interactions.

Another phenomenon that is intimately linked to resonant triad or higher-order interactions are extreme events that are amplified by nonlinearity \citep{sapsis2020statistics}. Rogue waves, for example, are characterized by their abnormal height and were linked to triad interactions both in deep \citep{drivas2016triad} and shallow water \citep{soomere2010rogue}. Similarly, \citet{farazmand2017variational} linked extreme dissipation events in turbulent Kolmogorov flow to a particular triad interaction that triggers fast energy transfer from large scales to the mean flow.

Because of their ubiquitous role turbulent and transitional flows, it comes as no surprise that triad interaction play an important role in reduced-order modeling. The class of models with dynamics that are restricted to triadic interactions that involve the mean flow, for example, is referred to as quasilinear models. Recent examples that use this specific type of triad truncation include the statistical state dynamics model by \cite{farrell2003structural} and the restricted nonlinear approximation by  \citet{thomas2014self,farrell2016statistical}. Rather than formally decomposing the flow into mean and fluctuations, the generalized quasilinear approximation by \cite{marston2016generalized} uses a spectral filter to separate different scales of motion. A modeling approach based on linear input-output dynamics is summarized in \cite{mckeon2017engine}. The latter reference contains a comprehensive overview of scale interactions in wall turbulence with emphasis on triad interactions. In the context of computational fluid mechanics, triad truncation can be understood as the wave-space manifestation of the turbulence closure problem. Large eddy simulation, for example, refers to the solution of the low-pass filtered Navier-Stokes equations and requires closure of the subgrid scale stress tensor. If large eddy simulation is conducted in wavenumber space, the closure problem explicitly becomes that of modeling the effect of triads that are effected by wavenumber cutoff on the resolved larger scales \citep[see, e.g.,][]{Pope2000}.

Higher-order statistical analysis refers to signal processing of time-series using higher-order spectra, or polyspectra. Of primary interest in the context of triadic interactions in self-excited and forced fluid flows is the bispectrum. Just like the power spectrum, the bispectrum can be estimated by ensemble-averaging products of realizations of the Fourier transform \citep{kim1979digital}. Unlike the power spectrum, which is real-valued by construction and carries no phase information, the bispectrum is capable of detecting quadratic phase-coupling by correlating different wave components. In early experimental work by \cite{lii1976bispectral}, the bispectrum was used to determine contributions of wavenumber triads to the energy transfer between different scales in atmospheric boundary-layer turbulence. The theoretical form of the bispectrum for Kraichnan's statistical models of homogenous turbulence, see above, has been devised and validated by \cite{herring1980theoretical,herring1992spectral}. Using the bispectrum, \cite{corke1989resonant} experimentally investigated the triadic interaction of phase-coupled input disturbances to a laminar boundary layer. Later, \cite{corke1991mode} studied the mode selection process and resonant phase locking in forced axisymmetric jets using the bispectrum. Other experimental studies that leverage the bispectrum to investigate triadic interactions and other nonlinear phenomena include the works of \cite{gee2010bicoherence} on the propagation of noise from a supersonic jet, by \cite{craig2019nonlinear} on second-mode instability in a hypersonic boundary layer, and by \cite{yamada2010two} on turbulence in plasmas. Also plasma physics, bispectral analysis earlier provided the first experimental evidence of an inverse energy cascade in drift-wave turbulence \citep{manz2008bispectral}. At somewhat larger scales, bispectral information significantly improved the accuracy of cosmological models that predict the spatial distribution of galaxies \citep{sefusatti2006cosmology}.

As large flow data of high temporal and spatial resolution have become ubiquitous, modal decomposition techniques are often applied as the primary means of flow analysis and data reduction. In fluid mechanics, the eigendecomposition of the covariance matrix is referred to as proper orthogonal decomposition \citep[POD,][]{Lumley:1970}. The resulting modes are orthonormal and optimally represent the data in terms of its variance. Classical space-only POD based on the spatial covariance matrix \citep{sirovich1987turbulence} is particularly well suited for low-order modeling \citep{aubry1988dynamics,deane1991low,noack2003hierarchy}, whereas frequency-domain, or spectral proper orthogonal decomposition (SPOD) is ideally suited to analyze statistically stationary data \citep{towneschmidtcolonius_2018_jfm,SchmidtColonius_2020_AIAAJ}. Dynamic mode decomposition \citep[DMD,][]{schmid2010dmd} permits analysis and modeling of flows in terms of the approximate eigendecomposition of the hypothetical evolution operator that maps the flow state from one snapshots to the next \citep[see also][]{rowley2009spectral}. 

Both POD and DMD are generally applicable to linear and nonlinear flow data, but neither method explicitly accounts for nonlinear interactions. The bispectrum, on the other hand, detects quadratic nonlinear interactions, but is only applicable to one-dimensional signals. To overcome these limitations in the context of stochastic estimation, \cite{baars2014proper} proposed a sequential approach in which SPOD of spatio-temporal data is followed by cross-bispectral analysis of the SPOD expansion coefficients. These coefficients, in turn, represent the dynamics of the structures that best represent the second-order statistics (variance or energy) of the data. In contrast to this sequential approach, the proposed framework directly computes structures that best represent the third-order statistics (skewness) of time- and space-resolved data.

%To overcome these limitations in the context of stochastic estimation, \cite{baars2014proper} serially combined higher-order statistical analysis and modal decomposition by computing the cross-bispectrum between SPOD expansion coefficients from spatially separated, synchronized microphone arrays. 

%In this work, bispectral mode decomposition is introduced as a direct means of computing modes that are associated with triadic interactions. 

The paper is organized as follows. We discuss higher-order spectra of time-series in \S\ref{sec:hos}. The theory of resonant triad interaction in laminar and turbulent flows is outlined in \S\ref{sec:threewave}. In \S\ref{sec:BMD}, bispectral mode decomposition is introduced, and some important symmetry properties are discussed in \S\ref{sec:symms}. In \S\ref{sec:test}, we first demonstrate the method on surrogate data with known phase coupling, before proceeding to analyze direct numerical simulation data of laminar cylinder flow at $\Rey=500$ in \S\ref{sec:cylinder}, particle image velocimetry data of turbulent flow over a flat plate at high angle of attack in \S\ref{sec:plate}, and large eddy simulation data of a transitional jet at $\Rey=3600$ in \S\ref{sec:jet}. Some implications for system identification and reduced-order modeling, and extension to higher-order statistics are briefly discussed in \S\ref{sec:discussion}. The main findings are summarized in \S \ref{sec:conclusions}. Details of the numerical algorithm and the convergence of the results are reported in appendices \ref{algorithm} and \ref{convergence}.

\sloppypar
The Matlab code used to compute the results is freely available. Two versions are provided. The first version solves the BMD based on the auto-bispectral density $S_{qqq}(f_1,f_2)$, as introduced in \S \ref{sec:BMD}. The second variant solves the corresponding problem for the cross-bispectral density $S_{qrs}(f_1,f_2)=\lim_{T\rightarrow\infty}\frac{1}{T}E\qty[\hat{q}(f_1)\hat{r}(f_2)\hat{s}(f_1+f_2)^*]$. This second variant was used to investigate azimuthal wavenumber interactions in \S \ref{sec:jet}, but can also be used to analyze interactions between different flow quantities.

\section{Background}
\label{sec:method}

\subsection{Higher-order spectra}\label{sec:hos}

We start by introducing the concept of the bispectrum for a stationary random signal $q(t)$ with zero mean, for which
\begin{equation}
R_q=E\qty[q(t)]=0,
\end{equation}
where $E\qty[\cdot]$ is the expectation operator and 
\begin{equation}\label{eqn:moments}
R_{q\dots q} = E\qty[q(t)q(t-\tau_1)q(t-\tau_2)\dots q(t-\tau_{n-1})]
\end{equation}
represents the $n$th-order moment of a stationary random signal. Through the introduction of the Fourier transform pair,
\begin{eqnarray}
\hat{q}(f) = \int_{-\infty}^{\infty}q(t)e^{-\ii2\pi f t}\dd t,\label{eqn:fft_cont_1} \\ q(t) = \int_{-\infty}^{\infty}\hat{q}(f)e^{\ii2\pi f t}\dd f, \label{eqn:fft_cont_2}
\end{eqnarray}
we can relate the signal’s power, or variance, 
\begin{equation}
E\qty[q(t)^2]=\int_{-\infty}^{\infty}S_{qq}(f)\dd f 
\end{equation}
to the power spectral density, or power spectrum,
\begin{equation}\label{eqn:psd}
S_{qq}(f) = \lim_{T\rightarrow\infty}\frac{1}{T}E\qty[\hat{q}(f)\hat{q}(f)^*].
\end{equation}
Note that the power spectrum is real, and therefore phase blind. It is directly related to the second-order moment, that is the autocorrelation function $R_{qq}(\tau)=E\qty[q(t),q(t-\tau)]$ via the Fourier transform,
\begin{equation}\label{eqn:wiener}
S_{qq}(f)=\int_{-\infty}^{\infty}R_{qq}(\tau)e^{-\ii2\pi f \tau}\dd \tau.%=\int_{-\infty}^{\infty}\Cmat_{qq}(f)e^{-\ii2\pi f \tau}\dd \tau
\end{equation}
This is the well-known Wiener–Khintchine theorem. Analogously, the bispectrum, or bispectral density, is defined as the double Fourier transform of the third moment,
\begin{equation}
S_{qqq}(f_1,f_2)=\int_{-\infty}^{\infty}\int_{-\infty}^{\infty}R_{qqq}(\tau_1,\tau_2)e^{-\ii2\pi (f_1\tau_1+f_2\tau_2)}\dd \tau_1 \dd \tau_2,
\end{equation}
and is a function of two frequencies, $f_1$ and $f_2$. Integration over the bispectrum recovers the expected value of the cubed signal, or skewness,
\begin{equation}
E\qty[q(t)^3]=\int_{-\infty}^{\infty}\int_{-\infty}^{\infty}S_{qqq}(f_1,f_2)\dd f_1 \dd f_2.
\end{equation}
Hence, the bispectrum decomposes the skewness of a stationary random signal into its frequency components. Since the skewness of symmetric distributions such as the Gaussian distribution is zero, the bispectrum is a direct measure of non-Gaussianity. Most important in the context of three-wave interactions, as discussed in \S \ref{sec:threewave} below, is the observation that the bispectrum correlates two frequency components to their sum. This can readily be seen from its definition in terms of the expectation operator,
\begin{equation}\label{eqn:Sqqq_cont}
S_{qqq}(f_1,f_2) = \lim_{T\rightarrow\infty}\frac{1}{T}E\qty[\hat{q}(f_1)^*\hat{q}(f_2)^*\hat{q}(f_1+f_2)],
\end{equation}
or, alternatively,
\begin{equation}\label{eqn:Sqqq_cont_alt}
S_{qqq}(f_1,f_2) = \lim_{T\rightarrow\infty}\frac{1}{T}E\qty[\hat{q}(f_1)\hat{q}(f_2)\hat{q}(f_1+f_2)^*].
\end{equation}
Equation (\ref{eqn:Sqqq_cont}) is consistent with definition (\ref{eqn:moments}) of the $n$th-order moment in terms of time delays $\tau$, whereas equation (\ref{eqn:Sqqq_cont_alt}) is associated with the interpretation of $\tau$ as time advances.

For further details on bi- and higher-order spectra, such as the definition of the bicoherence as a common normalization of the bispectrum, the reader is referred to the reviews by \cite{collis1998higher,brillinger1965introduction,kim1979digital,nikias1987bispectrum,nikias1993signal}. In \S \ref{sec:symms}, we will discuss the symmetry properties which bispectral mode decomposition inherits from the bispectrum.

\subsection{Triad interaction and resonance conditions}\label{sec:threewave}

Triad interaction is a specific type of three-wave coupling that results from quadratic nonlinearities in the governing equations.  Assume the dynamics of the state $\vb{q}=\vb{q}(\vb{x},t)$ are governed by an evolution equation of the form
\begin{equation} \label{eqn:dynsys}
\pdv{\vb{q}}{t}=\mathcal{L}\vb{q} + \mathcal{Q}(\vb{q},\vb{q}),
\end{equation}
where $\vb{x}=\qty[x,y,z]^T$ is the position vector, $\mathcal{L}$ a linear operator and $\mathcal{Q}(\cdot,\cdot)$ a quadratic (bilinear) nonlinearity. In the context of fluid flows governed by the incompressible Navier-Stokes equations (NSE), the convective term $\qty(\vb{u}\cdot\grad)\vb{u}$ represents such a quadratic nonlinearity. The vector $\vb{u}=\qty[u,v,w]^T$ contains the Cartesian velocity components. We further assume that there exists an an equilibrium solution $\vb{q}_0=\vb{q}_0(\vb{x})$ of equation (\ref{eqn:dynsys}) such that
\begin{eqnarray}\label{eqn:reynolds}
\mathcal{L}\vb{q}_0 + \mathcal{Q}(\vb{q}_0,\vb{q}_0)=0,
\end{eqnarray}
i.e., a stable, steady, laminar flow that is a solution of the NSE. Next, we decompose solutions to equation (\ref{eqn:dynsys}) into small but finite fluctuations around the equilibrium $\vb{q}_0$ in the form of a series
\begin{equation}\label{eqn:powerseries}
\vb{q}(\vb{x},t) = \vb{q}_0(\vb{x}) + \epsilon\vb{q}'(\vb{x},t) + \epsilon^2\vb{q}''(\vb{x},t) + \dots\quad\text{with} \quad 0<\epsilon \ll 1,
\end{equation}
of powers of $\epsilon$.
%$\vb{q}'$ are small (but finite) fluctuations around the equilibrium $\vb{q}_0$ which we express in the form of a power series
%\begin{equation}\label{eqn:powerseries}
%\vb{q}'(\vb{x},t;\epsilon) = \sum_{n=1}^\infty\epsilon^n\vb{q}'_n(\vb{x},t), \quad 0<\epsilon \ll 1.
%\end{equation}
Inserting equations (\ref{eqn:reynolds}) and (\ref{eqn:powerseries}) into equation (\ref{eqn:dynsys}), yields at leading order $\mathcal{O}(\epsilon)$ the equation
\begin{equation} \label{eqn:linsys_1}
\pdv{\vb{q}'}{t}=\mathcal{L}\vb{q}'.
\end{equation}
Due to linearity, periodic solutions to equation (\ref{eqn:linsys}) take the form
\begin{equation} \label{eqn:solnform}
\vb{q}'(\vb{x},t) \propto e^{\ii(\vb{k}\cdot\vb{x} - 2\pi f t)}.
\end{equation}
Combining equations (\ref{eqn:solnform}) and (\ref{eqn:linsys_1}) yield the dispersion relation
\begin{equation}\label{eqn:dispersion}
f=D(\vb{k})
\end{equation}
for the linear problem which relates frequency and the wavenumber vector. Suppose equation (\ref{eqn:linsys_1}) possesses $N$ periodic solutions
\begin{equation} \label{eqn:q1}
\vb{q}_n'(\vb{x},t) = A_n e^{\ii(\vb{k}_n\cdot\vb{x} - 2\pi f_n t)} + \text{c.c.},
\end{equation}
where $A=A(\vb{k},f)$ is a complex amplitude and c.c.~symbolizes the complex conjugate, $A^*e^{-\ii(\vb{k}\cdot\vb{x} - 2\pi f t)}$. For brevity, we introduce as
\begin{eqnarray}
\theta_n \equiv \vb{k}_n\cdot\vb{x} - 2\pi f_n t
\end{eqnarray}
the phase function. Also due to linearity, any sum of solutions $\vb{q}'_n$ also solves equation (\ref{eqn:linsys_1}). We may hence express the general solution as
\begin{equation} \label{eqn:q1sum}
\vb{q}'(\vb{x},t) = \sum_{n=1}^N \qty(A_n e^{\ii\theta_n} + \text{c.c.}).
\end{equation}
At $\mathcal{O}(\epsilon^2)$, we obtain the evolution equation obeyed by $\vb{q}''$,
\begin{equation} \label{eqn:dynsys_2}
\pdv{\vb{q}''}{t}=\mathcal{L}\vb{q}'' + \mathcal{Q}(\vb{q}',\vb{q}').
\end{equation}
In this equation, the nonlinearity first manifests in the form of the weakly nonlinear interaction of the linear solution $\vb{q}'$ with itself. Expanding the nonlinear term yields
%\begin{subequations}
\begin{eqnarray}\label{eqn:Qqq}
\mathcal{Q}(\vb{q}',\vb{q}') &=& \sum_{n=1}^N \qty(A_n e^{\ii\theta_n} + \text{c.c.})\sum_{m=1}^N \qty(A_m e^{\ii\theta_m} + \text{c.c.}) \\ \nonumber
&=& 2A_1A_1^* + A_1^2 e^{\ii2\theta_1} + A_1^{*2} e^{-\ii2\theta_1} + A_1A_2 e^{\ii(\theta_1+\theta_2)} \\ \nonumber && +A_1A_2^* e^{\ii(\theta_1-\theta_2)} +  A_1^*A_2 e^{\ii(-\theta_1+\theta_2)} + A_1^*A_2^* e^{\ii(-\theta_1-\theta_2)}\\ \nonumber&& + \dots.
\end{eqnarray}
%\end{subequations}
The first three terms of this sum result from the self-interaction of $\vb{q}_1'$. The first term has zero frequency and therefore contributes to the mean flow deformation. The second and third terms contribute to the first harmonic of $\vb{q}_1'$ which oscillates at $2f_1$. The fourth and fifth term are the sum- and difference-interactions of $\vb{q}_1'$ and $\vb{q}_2'$. Their respective phases are given as the sums and differences of their individual phases. In a slight change of notation for the subindices, denote by $\theta_j\equiv\theta_k\pm \theta_l$, or equivalently by
\begin{subequations}
\begin{eqnarray} \label{eqn:triadk} \vb{k}_j\equiv\vb{k}_k\pm \vb{k}_l,\\
f_j\equiv f_k\pm f_l,\label{eqn:triadf}
\end{eqnarray}
\end{subequations}
the sums and differences of any two wave components, $\theta_k=\vb{k}_k\cdot\vb{x} - 2\pi f_k t$ and $\theta_l=\vb{k}_l\cdot\vb{x} - 2\pi f_l t$, generated by quadratic interaction. Equations (\ref{eqn:triadk},b) are equivalent to equations (\ref{eqn:triad_wavenum},b). Now assume one of these newly generated wave components satisfies the dispersion relation for the linear system (\ref{eqn:dispersion}), that is,
\begin{equation}
f_j=D(\vb{k}_j).
\end{equation}
This has two important implications. First, the $\{f_j,\vb{k}_j\}$ component satisfies the linear portion of equation (\ref{eqn:dynsys_2}). Second, the linear portion of equation (\ref{eqn:dynsys_2}) is resonantly forced by the $\mathcal{O}(\epsilon^2)$ products in equation \eqn{eqn:Qqq} that share the same frequency and wavenumber. As a result, the corresponding wave component grows linearly in time and the system is said to be in resonance. Under these circumstances, equations \eqn{eqn:triadf} and  \eqn{eqn:triadk} establish a quadratic coupling of the phases between the $k$, $l$ and $k+l$ wave components. We refer to these frequency triplets $\{f_k,f_l,f_{k+l}\}$ or frequency index triplets $(k,l,k+l)$ as resonant (frequency) triads. Conversely, correlation between two wave components and their sum indicates the presence of a quadratic nonlinearity. For time signals, this property is exploited by the bispectrum, as can be seen from its definition, equation \eqn{eqn:Sqqq_cont}. The amplitude equations that govern the saturation of resonant waves are discussed in standard texts like \cite{craik1988wave} or \cite{schmid2001stability}.

For turbulent flows, equation (\ref{eqn:dynsys}), in general, does not possess a stable equilibrium solution $\vb{q}_0$. In this case, instead of fluctuations about an equilibrium, we may consider fluctuations $\vb{q}'$ around the mean flow
\begin{equation}
\bar{\vb{q}}(\vb{x})=\frac{1}{T}\int_0^\infty \vb{q}(\vb{x},t) \dd t.
\end{equation}
Inserting the (Reynolds) decomposition 
\begin{equation}\label{eqn:reydecomp}
\vb{q}(\vb{x},t)=\bar{\vb{q}}(\vb{x})+\vb{q}'(\vb{x},t)
\end{equation}
into equation (\ref{eqn:dynsys}) and averaging over time yields the mean flow equation,
\begin{equation} \label{eqn:rans}
\vb{0}=\mathcal{L}\bar{\vb{q}} + \mathcal{Q}(\bar{\vb{q}},\bar{\vb{q}}) + \overline{\mathcal{Q}(\vb{q}',\vb{q}')},
\end{equation}
where we used the property $\pdv{\bar{\vb{q}}}{t}=\overline{\pdv{\vb{q}'}{t}}=0$. These are the Reynolds-averaged Navier-Stokes (RANS) equations and $\overline{\mathcal{Q}(\vb{q}',\vb{q}')}$ are the Reynolds stresses. Linearizing about the mean flow and using equation (\ref{eqn:rans}) yields the equations obeyed by the fluctuations,
\begin{equation} \label{eqn:linsys}
\pdv{\vb{q}'}{t}=\mathcal{L}_{\bar{\vb{q}}}\vb{q}' - \overline{\mathcal{Q}(\vb{q}',\vb{q}')},
\end{equation}
where 
\begin{equation}\label{eqn:meanlinop}
\mathcal{L}_{\bar{\vb{q}}}\equiv \mathcal{L} + \mathcal{Q}(\cdot,\bar{\vb{q}}) + \mathcal{Q}(\bar{\vb{q}},\cdot)
\end{equation}
defines the linearized Navier-Stokes operator with respect to linearization about the mean flow. The Reynolds stresses $\overline{\mathcal{Q}(\vb{q}',\vb{q}')}$ in equations \eqn{eqn:rans} and \eqn{eqn:linsys} hinder the application of the resonant interaction theory presented above. To show that similar kinematic arguments apply to turbulent flows nevertheless, suppose that the fluctuating component $\vb{q}'(x,y,z,t)$ can be represented as a Fourier series,
\begin{equation}\label{eqn:fsexp_2}
	\vb{q}'(x,y,z,t) = \sum_{m,n=-\infty}^{\infty} \hat{\vb{q}}_{mn}(x,y) e^{\ii(k_m z - 2\pi f_n t)},
					 %&=& \bar{\vb{q}}(x,y) + \sum_{\substack{m,n=-\infty \\ (m,n)\neq (0,0)}}^{\infty} \hat{\vb{q}}_{mn}(x,y) e^{\ii(k_m z - 2\pi f_n t)}\label{eqn:fsexp},
\end{equation}
of periods $T$ in time and $L$ in the $z$-direction, where $f_n=n/T$ is frequency and $k_m=2\pi m/L$ wavenumber. The example of a flow with one homogeneous direction, $z$, and two inhomogeneous directions, $x$ and $y$, is choosen without loss of generality. Inserting equation \eqn{eqn:fsexp_2} into \eqn{eqn:reydecomp} yields the Reynolds decomposition of this flow,
\begin{equation}\label{eqn:reydecomp2}
\vb{q}(\vb{x},t)=\bar{\vb{q}}(x,y)+\sum_{m,n} \hat{\vb{q}}_{mn}(x,y) e^{\ii\theta_{mn}},
\end{equation}
where the mean is taken over time and in the $z$-direction.  For brevity, we denote as
\begin{eqnarray}
\theta_{mn}\equiv k_m z - 2\pi f_n t
\end{eqnarray}
the phase function and omit the limits of summation in equation \eqn{eqn:reydecomp2} and in the following. Due to the prior removal of the mean, the $(m,n)=(0,0)$ wave component does not contribute to the Fourier sum in equation \eqn{eqn:fsexp} since $\hat{\vb{q}}_{00}(x,y)=\vb{0}$. Inserting the Reynolds decomposition \eqn{eqn:reydecomp2} into the governing nonlinear equation \eqn{eqn:dynsys} yields
\begin{multline}
-2\pi\ii\sum_{m,n} f_n \hat{\vb{q}}_{mn} e^{\ii\theta_{mn}} = \mathcal{L}\bar{\vb{q}} + \mathcal{Q}(\bar{\vb{q}},\bar{\vb{q}}) + \sum_{m,n} \mathcal{L}\hat{\vb{q}}_{mn} e^{\ii\theta_{mn}} \\
    + \sum_{m,n} \qty[\mathcal{Q}(\bar{\vb{q}} ,\hat{\vb{q}}_{mn}) + \mathcal{Q}(\hat{\vb{q}}_{mn}, \bar{\vb{q}})]e^{\ii\theta_{mn}} 
	+ \sum_{m,n,p,q}  \mathcal{Q}(\hat{\vb{q}}_{mn}, \hat{\vb{q}}_{pq})e^{\ii\theta_{m+p\;n+q}}.
\label{eqn:fsexp}
\end{multline}
The generation of a $(m+p,n+q)$ wave component is apparent from the last term. This quadratic interaction process is analogous to the process described by equation \eqn{eqn:Qqq} and triads also take the form of equations \eqn{eqn:triadf} and \eqn{eqn:triadk}. The equations for the different Fourier components can be separated by exploiting the orthogonality of the complex exponential. Integration over $T$ and $L$, for example, isolates the zero-frequency and -wavenumber component,
\begin{equation}\label{eqn:zerocomp}
\vb{0} = \mathcal{L}\bar{\vb{q}} + \mathcal{Q}(\bar{\vb{q}},\bar{\vb{q}}) + \sum_{m,n}  \mathcal{Q}(\hat{\vb{q}}_{mn}, \hat{\vb{q}}_{-m-n}).%+ \sum_{\substack{m,n,p,q\\p=-m,q=-n}}  \mathcal{Q}(\hat{\vb{q}}_{mn}, \hat{\vb{q}}_{pq})e^{\ii\theta_{(m+p)(n+q)}}
\end{equation}
The sum in equation \eqn{eqn:zerocomp} comprises the zero-frequency and -wavenumber contributions of the quadruple sum in \eqn{eqn:fsexp}. These contributions results from the interaction of fluctuating wave components, $(m,n)$, with their conjugate counterparts, $(-m,-n)$. Comparing equations \eqn{eqn:zerocomp} and \eqn{eqn:rans} shows that this contribution corresponds to the Reynolds stresses. The equation for the $(m,n)$-th frequency-wavenumber component is isolated by multiplying by $e^{\ii\theta_{mn}}$ and integrating over $T$ and $L$ to obtain
\begin{multline}
-2\pi\ii f_n \hat{\vb{q}}_{mn} = \mathcal{L}\hat{\vb{q}}_{mn}
    + \mathcal{Q}(\bar{\vb{q}} ,\hat{\vb{q}}_{mn}) + \mathcal{Q}(\hat{\vb{q}}_{mn}, \bar{\vb{q}})
	+ \sum_{p,q}  \mathcal{Q}(\hat{\vb{q}}_{pq}, \hat{\vb{q}}_{m-p\;n-q}).    
	%+ \sum_{\substack{m',n',p,q\\m'+p=m,n'+q=n}}  \mathcal{Q}(\hat{\vb{q}}_{m'n'}, \hat{\vb{q}}_{pq})e^{\ii\theta_{m'+p\;n'+q}}
\label{eqn:mncomp}
\end{multline}
In terms of the linear operator defined in equation \eqn{eqn:meanlinop}, this equation reads
\begin{equation}\label{eqn:mncomp2}
-2\pi\ii f_n \hat{\vb{q}}_{mn} = \mathcal{L}_{\bar{\vb{q}}}\hat{\vb{q}}_{mn} + \sum_{p,q}  \mathcal{Q}(\hat{\vb{q}}_{pq}, \hat{\vb{q}}_{m-p\;n-q}).
\end{equation}
This form is similar to \eqn{eqn:rans} and similar arguments regarding the occurrence of resonances can be made. Equation \eqn{eqn:mncomp} illustrates how different wave components may contribute to the power spectral density at any given frequency, and also how contributions from different wave components to any given frequency can be identified by triple correlations of frequency components, i.e., by the bispectrum.

\section{Bispectral mode decomposition (BMD)}\label{sec:BMD}

The goal of this work is to devise a modal decomposition that reveals the presence of triadic nonlinear interactions from multidimensional data. As discussed in \S\S \ref{sec:hos} and \ref{sec:threewave}, quadratic phase coupling is characteristic of these interactions and can be detected by the bispectrum. To compute modes that exhibit quadratic phase coupling over extended portions of the flow field, we require the decomposition to optimally represent the data in terms of an integral measure of the bispectral density. BMD may be understood as the extension of the analysis of time signals using higher-order spectra to multidimensional datasets, or \emph{vice versa}, as an extension of spectral proper orthogonal decomposition to higher-order spectra. In particular, consider data that is given as a series of $N_t$ consecutive flow fields 
\begin{equation}
\vb{q}(\vb{x},t_j)\in \mathbb{C}^{M\times 1}, \;j=1,2,\dots,{N_t},
\end{equation}
that are evenly spaced in time. Let $M=N_{\text{vars}}N_xN_yN_z$ be the number of spatial degrees of freedom per time instant, with $N_{\text{vars}}$ as the number of variables in the state vector, and $N_x,\;N_y,\;N_z$ the numbers of grid points in the Cartesian directions, respectively. The bispectal density is defined in equation \eqn{eqn:Sqqq_cont} as the expected value of the product of two frequency components with their sum. As an estimator for the bispectrum we adapt Welch's method \citep{welch1967use}, which is an asymptotically consistent spectral estimator for the power spectral density. Welch's method is based on the ergodicity hypothesis. It assumes that the time average  in equation \eqn{eqn:psd} can be estimated by an ensemble average over a number of $\Nblk$ realizations of the Fourier transform. The underlying assumption is that the time series $\vb{q}(\vb{x},t_j)$ is statistically stationary. Each realization is obtained as the discrete-time Fourier transform of one of $\Nblk$ segments consisting of $\Nfft$ snapshots. To decrease the variance of the estimate, the number of segments is inflated by allowing consecutive segments to overlap by $\Novlp$ elements. Given a total number of $N_t$ snapshots, we obtain a number of 
\begin{equation}
	\Nblk=\text{floor}\qty(\frac{N_t-\Novlp}{\Nfft-\Novlp})
\end{equation} 
	realizations of the Fourier transform, $\hat{\vb{q}}^{[1]}(\vb{x},f),\hat{\vb{q}}^{[2]}(\vb{x},f),\dots\hat{\vb{q}}^{[\Nblk]}(\vb{x},f)$.
The discrete Fourier transform and its inverse are defined as
    \begin{eqnarray}
        \hat{\vb{q}}(\vb{x},f_k) = \sum_{j=0}^{\Nfft-1}\vb{q}(\vb{x},t_{j+1})\ee^{-\ii2\pi jk/\Nfft}, \quad  k=0,\dots,\Nfft-1, \;\text{and}\label{eqn:fft_disc_1}\\
        \vb{q}(\vb{x},t_{j+1}) = \frac{1}{\Nfft}\sum_{k=0}^{\Nfft-1}\hat{\vb{q}}(\vb{x},f_k)\ee^{\ii2\pi jk/\Nfft}, \quad  j=0,\dots,\Nfft-1. \label{eqn:fft_disc_2}
    \end{eqnarray}
The time step $\Delta t$ between consecutive snapshots determines the sampling frequency $f_s=1/\Delta t$ and thereby the  Nyquist frequency $f_N=f_s/2$. Computation of the bispectrum further requires the product of two frequency components. To compute products of Fourier coefficients of multidimensional data, we use the entry-wise, or Hadamard product defined as $\qty(\vb{A}\circ\vb{B})_{jk}=\vb{A}_{jk}\vb{B}_{jk}$. It applies to two matrices $\vb{A}$ and $\vb{B}$ of the same dimensions. For brevity, we introduce the shorthands
    \begin{gather}
    \hat{\vb{q}}_k \equiv \hat{\vb{q}}(\vb{x},f_k), \;\text{and} \\
    \hat{\vb{q}}_{k\circ l} \equiv \hat{\vb{q}}(\vb{x},f_k)\circ\hat{\vb{q}}(\vb{x},f_l),
    \end{gather}
for the $k$-th frequency component of the discrete-time Fourier transform and the spatial entry-wise product of two realizations of the Fourier transform at frequencies $f_k$ and $f_l$, respectively. At the heart of bispectral mode decomposition is the definition of an integral measure 
\begin{equation}\label{eqn:b}
		b(f_k,f_l)\equiv E\qty[\int_\Omega\hat{\vb{q}}_{k}^*\circ\hat{\vb{q}}_{l}^*\circ\hat{\vb{q}}_{k+l}\;\dd\vb{x}] = E\qty[{\hat{\vb{q}}_{k\circ l}}^H\vb{W}\hat{\vb{q}}_{k+l}] = E\qty[\expval{\hat{\vb{q}}_{k\circ l},\hat{\vb{q}}_{k+l}}], %S_{qqq}(f_1,f_2) = \lim_{T\rightarrow\infty}\frac{1}{T}E\qty[\hat{q}(f_1)\hat{q}(f_2)\hat{q}(f_1+f_2)^*].
	\end{equation}
of the point-wise bispectral density. By $(\cdot)^*$, $(\cdot)^T$, and $(\cdot)^H$ we distinguish the scalar complex conjugate, transpose, and complex transpose, respectively. $\vb{W}$ is the diagonal matrix of spatial quadrature weights and $\Omega$ the spatial domain over which the flow is defined. The weighted inner product, 
\begin{equation}
\expval{\vb{q}_1,\vb{q}_2}=\vb{q}_1^H\vb{W}\vb{q}_2,
\end{equation}
is introduced as the discrete analogue to spatial integration. In equation \eqn{eqn:b}, the combination of the $l$-th and $k$-th frequency components into ${\hat{\vb{q}}_{k\circ l}}^H\equiv\hat{\vb{q}}_{k}^*\circ\hat{\vb{q}}_{l}^*$ is merely notational.
In the following, however, we explicitly take into account the causal relation between the sum-frequency component $\hat{{q}}_{k+l}$ (effect), and the product of the $l$-th and $k$-th frequency components, $\mathcal{Q}(\hat{{q}}_{k}, \hat{{q}}_{l})\propto \hat{{q}}_{k} \hat{{q}}_{l}$ (cause), that form a resonant triad, and define two linear expansions
    \begin{eqnarray}
        \vb*{\phi}_{k\circ l}^{[i]}(\vb{x},f_k,f_l) 
        &=& \sum_{j=1}^{\Nblk} a_{ij}(f_{k+l})\hat{\vb{q}}_{k\circ l}^{[j]}\label{eqn:exp1}\quad\text{(cross-frequency field)}, \\ 
    	\vb*{\phi}_{k+ l}^{[i]}(\vb{x},f_{k+l}) &=& \sum_{j=1}^{\Nblk} a_{ij}(f_{k+l})\hat{\vb{q}}_{k+l}^{[j]}\label{eqn:exp2}\quad\text{(bispectral modes)},
    \end{eqnarray}
that share a common set of expansion coefficients $a_{ij}$. In the light of equation \eqn{eqn:dynsys}, this corresponds to discriminating between the resonantly forced wave component at $\mathcal{O}(\epsilon^2)$ and the product of the two interacting components of $\mathcal{O}(\epsilon)$. Equations \eqn{eqn:exp1} and \eqn{eqn:exp2} describe expansions into the spaces spanned by the ensembles of realizations of $\hat{\vb{q}}_{k\circ l}$ and $\hat{\vb{q}}_{k+l}$, respectively. We will refer to $\vb*{\phi}_{k+l}$ as \emph{bispectral modes}. Bispectral modes are linear combinations of Fourier modes and can be interpreted as observable physical structures. The multiplicative cross-frequency fields $\vb*{\phi}_{k\circ l}$, on the contrary, are maps of phase-alignment between two frequency components that may not directly be observed. A more compact form of equations \eqn{eqn:exp1} and \eqn{eqn:exp2} is 
    \begin{eqnarray} \label{eqn:phidef}
        \vb*{\phi}_{k\circ l}^{[i]} &=& \hat{\vb{Q}}_{k\circ l}\vb{a}_i,\\
    	\vb*{\phi}_{k+l}^{[i]}  &=& \hat{\vb{Q}}_{k+l}\vb{a}_i,
    \end{eqnarray}
where $\vb{a}_i = \qty[a_{i1}(f_{k+l}),\;a_{i2}(f_{k+l}),\;\dots,\;a_{i\Nblk}(f_{k+l})]^T$ denotes the $i$-th vector of expansion coefficients for the $(k,l)$ frequency doublet, and $\hat{\vb{Q}}_{k\circ l},\hat{\vb{Q}}_{k+l}\in \mathbb{C}^{M\times \Nblk}$ are the data matrices
    \begin{equation}\label{eqn:data_mats}
        \hat{\vb{Q}}_{k\circ l} \equiv \mqty[
        \vertbar     & \vertbar     &       & \vertbar      \\
        \hat{\vb{q}}_{k\circ l}^{[1]} & \hat{\vb{q}}_{k\circ l}^{[2]} & \cdots & \hat{\vb{q}}_{k\circ l}^{[\Nblk]}  \\
        \vertbar     & \vertbar     &       & \vertbar] , \;
%    \end{equation}
%    \begin{equation}
        \hat{\vb{Q}}_{k+l} \equiv \mqty[
        \vertbar     & \vertbar     &       & \vertbar      \\
        \hat{\vb{q}}_{k+l}^{[1]} & \hat{\vb{q}}_{k+l}^{[2]} & \cdots & \hat{\vb{q}}_{k+l}^{[\Nblk]}  \\
        \vertbar     & \vertbar     &       & \vertbar].
    \end{equation}
The goal of bispectral mode decomposition is to compute modes that optimally represent the data in terms of the integral bispectral density. That is, we seek the set of expansion coefficients $\vb{a}_1$ that maximizes the absolute value of $b(f_k,f_l)$ as defined in equation \eqn{eqn:b}.  To guarantee boundedness of the expansion, we require the coefficient vector to be a unit vector with $\|\vb{a}_1\|=1$. The optimal $\vb{a}_1$ hence  has to satisfy
\begin{eqnarray}\label{eqn:a1}
      \vb{a}_1=\argmax_{\|{\vb{a}}\|=1}\qty|E\left[{\vb*{\phi}^{[1]H}_{k\circ l}}\vb{W}\vb*{\phi}^{[1]}_{k+l}\right]| \nonumber
	= \argmax_{\|{\vb{a}}\|=1}\qty|E\left[\vb{a}^H{\hat{\vb{Q}}}^H_{k\circ l}\vb{W}\hat{\vb{Q}}_{k+l}\vb{a}\right]| \\
	= \argmax\qty|\frac{\vb{a}^H E\left[{\hat{\vb{Q}}}^H_{k\circ l}\vb{W}\hat{\vb{Q}}_{k+l}\right]\vb{a}}{\vb{a}^H\vb{a}}|
	= \argmax\qty|\frac{\vb{a}^H\vb{B}\vb{a}}{\vb{a}^H\vb{a}}|,
\end{eqnarray}
	where we introduced 
\begin{equation}\label{eqn:bsd}
    \vb{B}=\vb{B}(\vb{x},\vb{x}',f_k,f_l)\equiv \frac{1}{\Nblk}\hat{\vb{Q}}_{k\circ l}^H\vb{W}\hat{\vb{Q}}_{k+l}
\end{equation}
    as the weighted (auto-) bispectral density matrix. Here, `auto' implies that three frequency components of the same state $\vb{q}$ are involved. Consistent with the nomenclature for higher-order statistics of time-signals, we reserve the term cross-bispectrum for third-order statistics such as $S_{qqr}(f_1,f_2)$ or $S_{qrs}(f_1,f_2)$ that involve two or three different fields. In the context of multidimensional data, spatial cross-correlation between any two locations $\vb{x}$ and $\vb{x}'$ is always implied. The final expression in equation (\ref{eqn:a1}) corresponds to finding the complex vector $\vb{a}\in\mathbb{C}^{\Nblk\times1}$ that maximizes the absolute value of the Rayleigh quotient of the complex, non-Hermitian, square matrix $\vb{B} \in \mathbb{C}^{\Nblk\times\Nblk}$. This problem is directly related to the numerical range, or field of values, which is defined as the set of all Rayleigh quotients of a matrix,
	\begin{equation}\label{eqn:numrange}
		F(\vb{B}) = \qty{\frac{\vb{a}^H\vb{B}\vb{a}}{\vb{a}^H\vb{a}} : \vb{a}\in\mathbb{C}^{\Nblk\times1}, \vb{a}\neq\vb{0}}.
	\end{equation}
    The largest absolute value the numerical range can attain defines the numerical radius
	\begin{equation}\label{eqn:numradius}
		r(\vb{B}) = \max \qty{|\lambda| : \lambda\in F(\vb{B})}.
	\end{equation}	    
   	Therefore, the maximization problem (\ref{eqn:a1}) is equivalent to finding the vector $\vb{a}_1$ associated with the numerical radius of $\vb{B}$,
	\[
	r(\vb{B}) = \max\qty|\frac{\vb{a}_1^H\vb{B}\vb{a}_1}{\vb{a}_1^H\vb{a}_1}|.
	\]
	Geometrically, the numerical radius can be interpreted as the radius of the smallest circle about the origin that contains the field of values. It can be shown \citep[see, e.g.,][]{horn1991topics,watson1996computing} that the numerical radius corresponds to the largest eigenvalue $\lambda_{\max}$ that the Hermitian matrix    
	\begin{equation}\label{eqn:H}
   \vb{H}(\theta) = \frac{1}{2}\qty(\ee^{\ii\theta}\vb{B}+\ee^{-\ii\theta}\vb{B}^H)
   \end{equation}
	can attain for some angle $0\leq\theta<2\pi$, i.e.,
	\begin{equation} \label{eqn:maxtheta}
	 	r(\vb{B}) = \max_{0\leq\theta<2\pi} \lambda_{\max}\qty(\vb{H}(\theta)).%=\max_{0\leq\theta<\pi} \lambda_{\text{maxabs}}\qty(\vb{H}(\theta))
	\end{equation}	
	Denote by $\theta_{1}$ the angle for which expression (\ref{eqn:maxtheta}) assumes its maximum value and by $\vb{a}_1$ the leading eigenvector such that
	\[
	\vb{H}(\theta_{1})\vb{a}_1 = \lambda_1\vb{a}_1.
	\]	
	Then $\lambda_1=\lambda_{\max}\qty(\vb{H}(\theta_{1}))$ is the numerical radius and $\vb{a}_1$ maximizes the absolute value of the Rayleigh quotient of $\vb{B}$, i.e.,
	\[
	r(\vb{B}) = \lambda_1 = \qty|\frac{\vb{a}_1^H\vb{B}\vb{a}_1}{\vb{a}_1^H\vb{a}_1}|.
	\]
	To distinguish $\lambda_1$ from the traditional definition of the bispectrum for time-series, we will refer to
	\begin{equation}
	\label{eqn:lambda1}
	\lambda_1(f_k,f_l) \quad\text{(complex mode bispectrum)}
	\end{equation}
	as the \emph{complex mode bispectrum}. It is tempting to approximate the eigenpair $(\lambda_1,\vb{a}_1)$ by solving equation (\ref{eqn:maxtheta}) over a discretized interval $\theta\in(0,2\pi]$ to find $\theta_1$. Instead of this brute-force approach, we employ the much more elegant and efficient algorithm by \cite{he1997algorithm}. The algorithm is reproduced, with minor modifications, in appendix \ref{algorithm}.

\subsection{Symmetries and regions of the bispectrum}\label{sec:symms}

\begin{figure}
\begin{center}
  \includegraphics[trim={12cm 8.5cm 11.8cm 7.5cm},clip,width=0.5\textwidth]{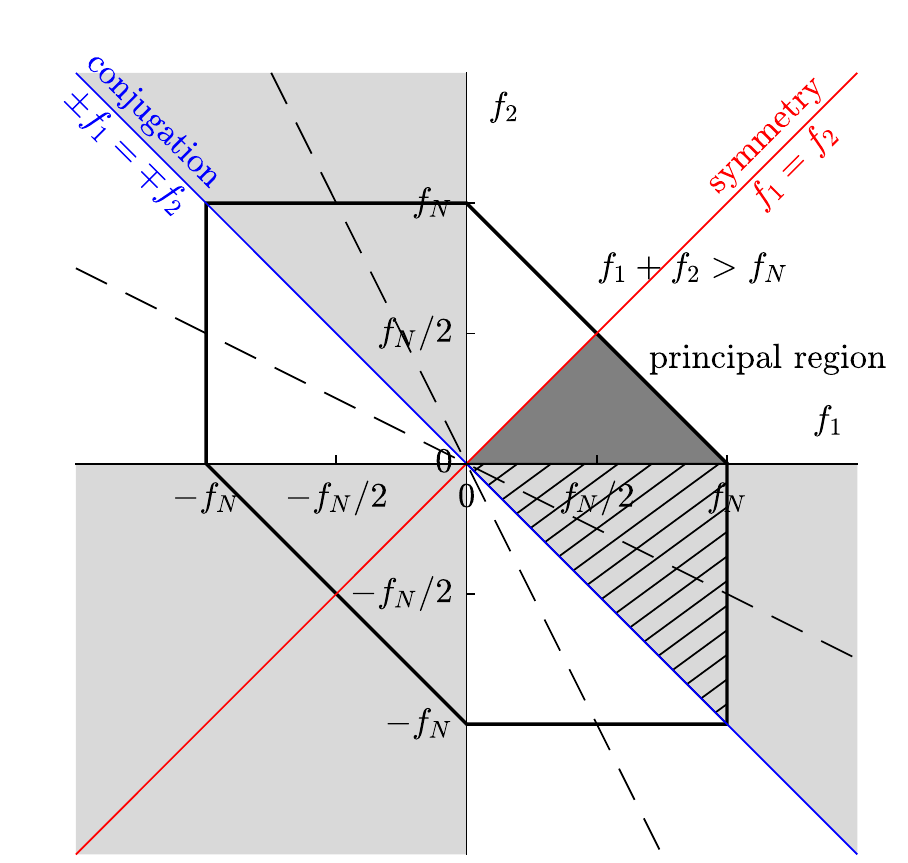}
  \end{center}
\caption{Symmetry regions of the bispectrum and mode bispectrum. The dark shaded triangle indicates the principal region of non-redundant information of the classical bispectrum. White regions inside the hexagon contain the same information as the principal region. Light gray shading indicates complex conjugation. The principal region of the bispectrum corresponds to sum-interactions with $f_3=f_1+f_2$. The hatched segment corresponds to difference-interactions with $f_3=f_1-f_2$. Both sum- and difference-interactions can be analyzed using BMD.}
\label{fig:symm}       % Give a unique label
\end{figure}
\subsubsection{Temporal homogeneity} The Nyquist frequency limit restricts the discrete bispectrum to the hexagonal region outlined in figure \ref{fig:symm}. For the (auto-) bispectrum of a time signal, defined by equation \eqn{eqn:Sqqq_cont}, it suffices to compute the principal region $0\leq f_2 \leq f_N/2$ and $f_2 \leq f_1 \leq f_N-f_2$. The remaining 11 regions then carry the same information as the principal region and its complex conjugate. This symmetry of the bispectrum results from the symmetry of the discrete-time Fourier transform for time stationary signals, which translates into, among others, the following symmetries for the bispectrum for signals: $S_{qqq}(f_1,f_2)=S_{qqq}(f_2,f_1)=S_{qqq}^*(-f_1,f_2)=S_{qqq}^*(-f_2,f_1+f_2)=S_{qqq}^*(-f_1,f_1+f_2)$. In particular, this implies that sum- and difference-interactions in equation (\ref{eqn:triadf}) do not have to be considered separately.

The mode bispectrum defined by equation (\ref{eqn:lambda1}), on the other hand, distinguishes between sum- and difference-interactions. Take as examples two triads involving the same two frequencies, $f_1$ and $f_2$. Let the first triad be the sum-interaction $\{f_1,f_2,f_1+f_2\}$, and the second the difference-interaction $\{f_1+f_2,-f_1,f_2\}$. By exploiting the symmetry of the Fourier transform, which implies that $\hat{q}(-f_1)=\hat{q}(f_1)$, and the commutativity of the three factors in bispectrum, we may readily show that $S_{qqq}(f_1,f_2)=S_{qqq}^*(f_1+f_2,-f_1)$. The computation of the complex mode bispectrum, on the contrary, is based on the bispectral density matrix, $\vb{B}$, which differentiates between sum- and difference-interactions. In particular, we have
\begin{eqnarray*}\label{eqn:bsd_ineq}
    \hat{\vb{Q}}_{k\circ l}^H\vb{W}\hat{\vb{Q}}_{k+l}  &\neq& (\hat{\vb{Q}}_{k+l\circ -k}^H\vb{W}\hat{\vb{Q}}_{l})^*, \text{and hence } \\ \vb{B}(\vb{x},\vb{x}',f_k,f_l) &\neq& \vb{B}^*(\vb{x},\vb{x}',f_{k+l},f_{-k}),
\end{eqnarray*}
in general. This loss of commutativity results from the causal relationship established by equations \eqn{eqn:exp1} and \eqn{eqn:exp2}. It is also apparent that the bispectral modes associated with the higher frequency of the sum-interaction, $f_1+f_2$, and the lower frequency of the difference-interaction, $f_2$, must be a linear combinations of the corresponding Fourier modes, $\hat{\vb{q}}^{[j]}_{k+l}$ and $\hat{\vb{q}}^{[j]}_{l}$, respectively. The symmetry relations of the bispectrum and mode bispectrum are summarized schematically in figure \ref{fig:symm}.

\subsubsection{Spatial homogeneity}\label{sec:space_hom}
Taking spatial homogeneity into account is beneficial not only in terms of computational efficiency, but also for convergence of the spectral estimate and interpretability. Spatial symmetries, such as periodicity, are accounted for through discrete-space Fourier transformation in the corresponding directions. The transformation to wavenumber space permits the identification of the phase coupling between different spatial scales in the same way as the temporal transform for time scales. Take as an example data that are invariant under translation in the $x$ direction. Analogous to the treatment of time for a stationary random signal, we may assume that the spatial bicorrelation in $x$ only depends on the relative distances $x-\xi_1$ and $x-\xi_2$. We may hence define the spatio-temporal bicorrelation as
\begin{equation}\label{eqn:Rqqq_spacetime}
R_{qqq}(\xi_1,\xi_2,\tau_1,\tau_2)=E\qty[q(x,t),q(x-\xi_1,t-\tau_1),q(x-\xi_1,t-\tau_2)].
\end{equation}
The spatio-temporal Fourier transform of $R_{qqq}(\xi_1,\xi_2,\tau_1,\tau_2)$ yields the spatio-temporal bispectrum
\begin{equation}\label{eqn:Sqqq_spacetime}
S_{qqq}(k_1,k_2,f_1,f_2) = \lim_{T\rightarrow\infty}\frac{1}{T}E\qty[\hat{q}(k_1,f_1)^*\hat{q}(k_2,f_2)^*\hat{q}(k_1+k_2,f_1+f_2)],
\end{equation}
where we denote as $k$ the $x$-component of the wavenumber vector. Equation \eqn{eqn:Sqqq_spacetime} is analogous to \citet{yamada2010two}'s definition of the two-dimensional bispectrum. Equation (\ref{eqn:Sqqq_spacetime}) implies that the bispectrum is to be computed for each wavenumber doublet $\{k_1,k_2\}$ individually. Triadically-consistent wavenumber triplets $\{k_1,k_2,k_1+k_2\}$ are referred to as spatial triads. For the example of three-dimensional data that are translationally invariant in the $x$-direction, the BMD is therefore computed from the discrete-space discrete-time transformed data $\hat{\vb{q}}(\vb{z},k,f)$, where by $\vb{z}=\qty[y,z]^T$ we denote the position vector of the remaining inhomogeneous directions. Accordingly, the bispectral density matrix specializes to
\begin{equation}\label{eqn:homdir}
\vb{B}(\vb{x},\vb{x}',f_k,f_l) \rightarrow \vb{B}(\vb{z},\vb{z}',k_i,k_j,f_k,f_l),
\end{equation}
and is computed from $\hat{\vb{q}}(\vb{z},k_i,f_k)$, $\hat{\vb{q}}(\vb{z},k_j,f_l)$, and $\hat{\vb{q}}(\vb{z},k_{i+j},f_{k+l})$. The resulting BMD modes are two-dimensional and the spatio-temporal mode bispectrum is defined in the four-dimensional wavenumber-wavenumber-frequency-frequency domain. For doubly-homogeneous flows such as pipe or Couette flows, the BMD is computed for all combinations between two wavenumbers and frequency and the resulting modes are one-dimensional.

\subsubsection{Regions of the bispectrum}

\begin{figure}
  \includegraphics[trim={2.3cm 0cm 2.8cm 0cm},clip,width=0.5\textwidth]{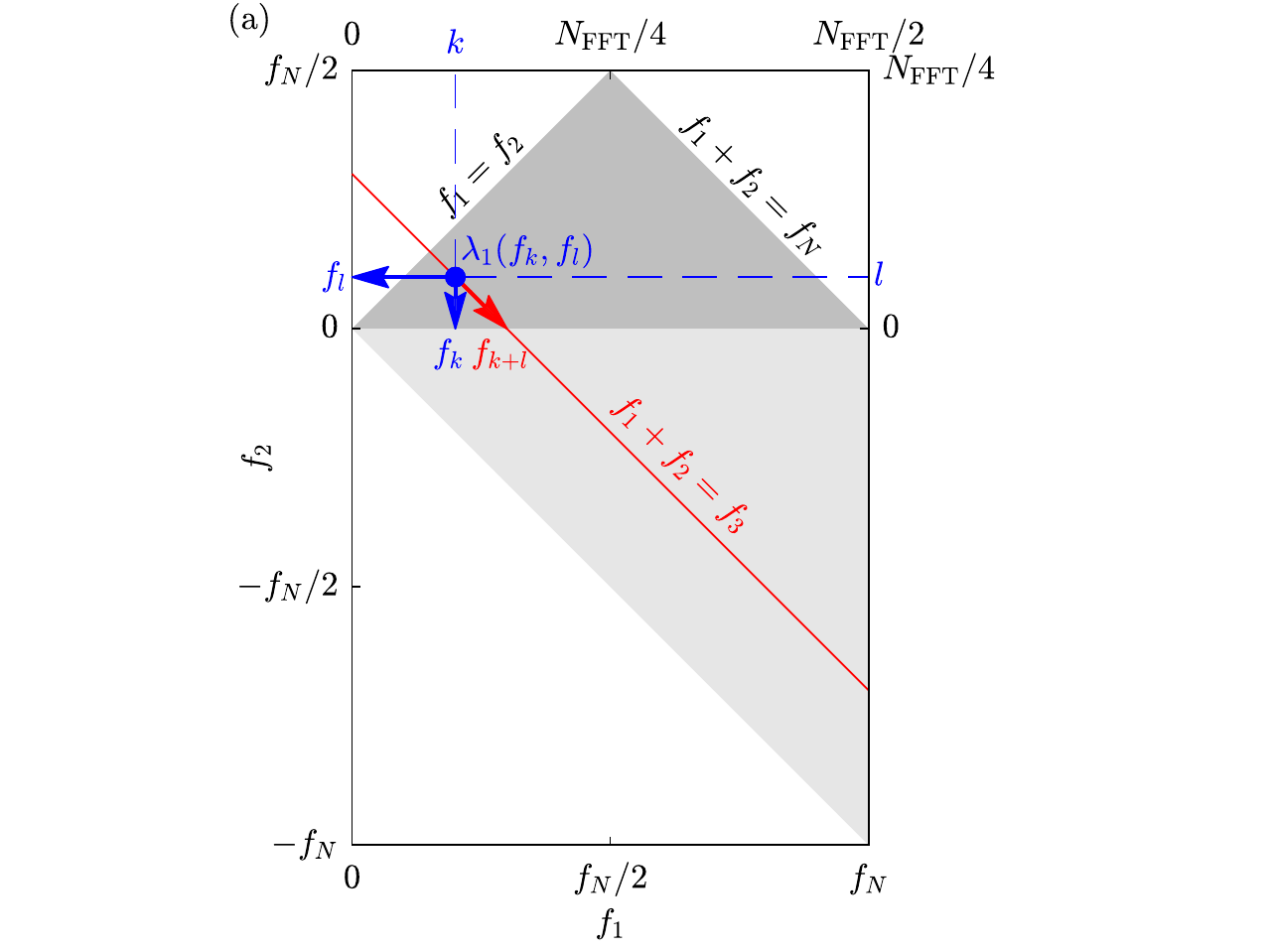}\\
  \includegraphics[trim={2.3cm 0cm 2.8cm 0cm},clip,width=0.5\textwidth]{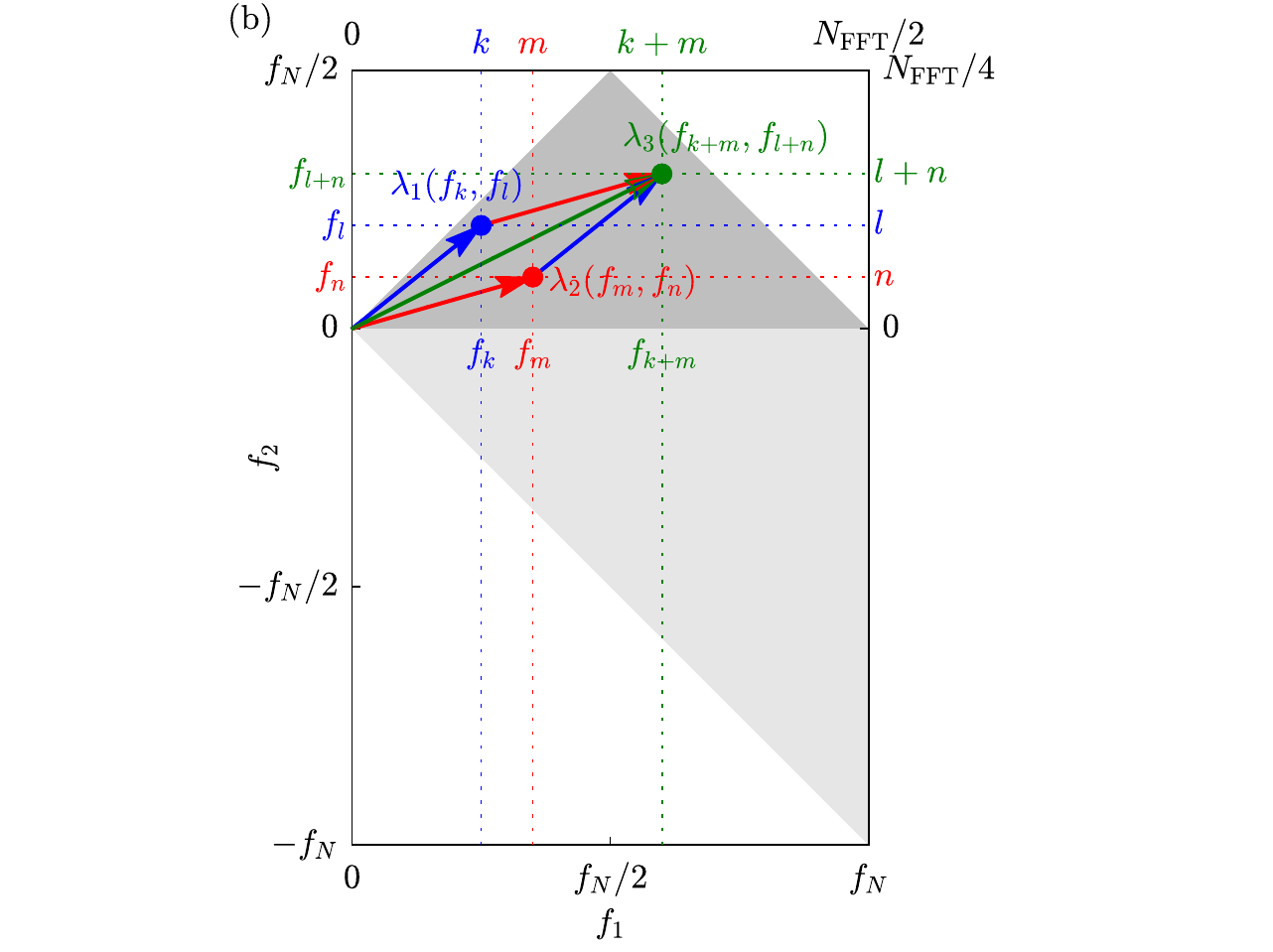}
\caption{Frequency triads in the mode bispectrum: (a) a local maximum at $\lambda_1(f_k,f_l)$ indicates phase-coupling between the components of the triad $\{f_k,f_l,f_{k+l}\}$ due to quadratic nonlinearity; (b) the sum components generated by two triads, $\{f_k,f_l,f_{k+l}\}$ and $\{f_m,f_n,f_{m+n}\}$, can interact to generate a new wave component $\{f_{k+m},f_{l+n}\}$. Harmonics and the mean flow distortion are generated by sum and difference self-interactions of the forms $\{f_k,f_k,f_{2k}\}$ and $\{f_k,f_{-k},0\}$, respectively.}
\label{fig:principal}       % Give a unique label
\end{figure}
 
Figure \ref{fig:principal} shows a schematic of the sum and difference regions of the mode bispectrum. All other regions contain redundant information or lie outside of the Nyquist limit. It hence suffices to plot this region. For brevity, we will often use integer frequency index doublets $(k,l)$,  or triplets $(k,l,k+l)$, instead of frequency doublets $\{f_k,f_l\}$, or triplets $\{f_k,f_l,f_{k+l}\}$, to represent triads. Figure \ref{fig:principal}(b) illustrates the generation of new wave components through triad interactions. In the examples discussed in \S \ref{sec:examples}, we will observe that this mechanism, starting from the self-interaction of a self-excited or forced fundamental mode, often leads to a distinct grid pattern of the bispectrum.

\subsection{Derived quantities}\label{sec:derived}

\subsubsection{Summed mode bispectrum}\label{sec:sumspec}
Since the mode bispectrum is complex, we visualize its modulus, the \emph{(magnitude) mode bispectrum} $|\lambda_1(f_k,f_l)|$, and argunent, the \emph{phase mode bispectrum} $\arg(\lambda_1(f_k,f_l))$, separately. A detailed discussion of the properties of phase of the classical bispectrum was provided by \cite{kim1980bispectrum}. Analogous to the common definition of a summed bispectrum for time signals, we furthermore define the \emph{summed mode spectrum} as
\begin{gather}
\Lambda_1(f)\equiv\frac{1}{N(f)}\sum_{f=f_1+f_2}|\lambda_1(f_1,f_2)|\quad\text{(summed mode spectrum)},
\end{gather}
where $N(f)$ is the number of frequency doublets $\{f_1,f_2\}$ that contribute to any frequency $f=f_1+f_2$, that is, the number of terms in the sum. Graphically, this corresponds to summing $\lambda_1$ along diagonals of slope $-1$, i.e., lines of constant frequency in the mode bispectrum. Take as an example the red line of constant frequency $f_3$ in figure \ref{fig:principal}. Peaks in the summed mode spectrum indicate that the corresponding frequencies are involved in quadratic nonlinear interactions, but without discriminating between the contributing triads.

\subsubsection{Interaction maps}\label{sec:intermaps}
Equations (\ref{eqn:a1}) implies that the mode bispectrum derives from the spatial integration of the Hadamard product \begin{equation}\label{eqn:interaction_map}
        \vb*{\psi}_{k,l}(\vb{x},f_k,f_l)\equiv\qty|\vb*{\phi}_{k\circ l}\circ\vb*{\phi}_{k+l}|\quad\text{(interaction map)}.
\end{equation}
We hence may interpret the field $\vb*{\psi}_{k,l}$ as an \emph{interaction map} that quantifies the average local bicorrelation between the three frequency components $f_l$, $f_k$, and $f_k+f_l$ involved in a triad. The interaction map augments the bispectral modes in that it indicates regions of activity of triadic interaction.

\subsection{Hypothesis testing}\label{sec:test}

Recall that the goal of BMD is to identify flow structures associated with frequency triads for which the zero-sum condition, equation \eqn{eqn:triad}, holds and that the indicator for this condition is a maximum in the mode bispectrum, $|\lambda_1(f_1,f_2)|$. For the outcomes of BMD to be interpretable, we furthermore require the method to reject (ideally also in the presence of noise) arbitrary non-resonant frequency triplets and higher-order wave interactions. The hypothesis that summarizes these requirements is that the mode bispectrum, (i), indicates triadically interacting wave components with $f_1\pm f_2\pm f_3=0$, while, (ii), rejecting triplets with $f_1\pm f_2\pm f_3\neq0$ and, (iii), other $N$-wave interactions with $f_1\pm f_2\pm f_3\pm\dots\pm f_{N}=0$. Since the nonlinearities in most physical systems are limited to cubic order at most, we restrict this test to $N=4$. To isolate and test these three aspects, we generate surrogate data consisting of a superposition of a certain number of $N_\text{peaks}$ waves with specified phase-coupling. The $i$-the realization of the data is generated as
\begin{equation}
\vb{q}^{[i]}(x,t)=\sum_{j=1}^{N_\text{peaks}}A_j \cos(k_{j}x-2\pi f_j t+\theta_0^{[i]}),
\end{equation}
where $A_j$ are the wave amplitudes, $k_j$ the wavenumbers in $x$, and $\theta_0^{[i]}$ a phase offset that is used to randomize the phase between realizations. We consider waves of unit amplitude and random wavenumbers $k\in[0,5]$ that evolve in a 1-D spatial domain $x\in[0,2\pi]$, which is discretized by 100 equidistant points. A total number of 1280 snapshots, separated in time by $\Delta t=1$, are segmented onto 10 blocks of length $\Nfft=128$ each. Refer to table \ref{tab:data} for a summary of the spectral estimation parameters. Neither frequencies nor wavenumbers can, in general, be expected to be harmonic multiples of their respective domains. We therefore deliberately select frequencies that do not have this property and randomize the wavenumbers. Following best practices in spectral estimation, a standard Hann window is applied to each block to reduce spectral leakage. 

\begin{figure}
  \includegraphics[trim={0cm 0cm 0cm 0cm},clip,width=1\textwidth]{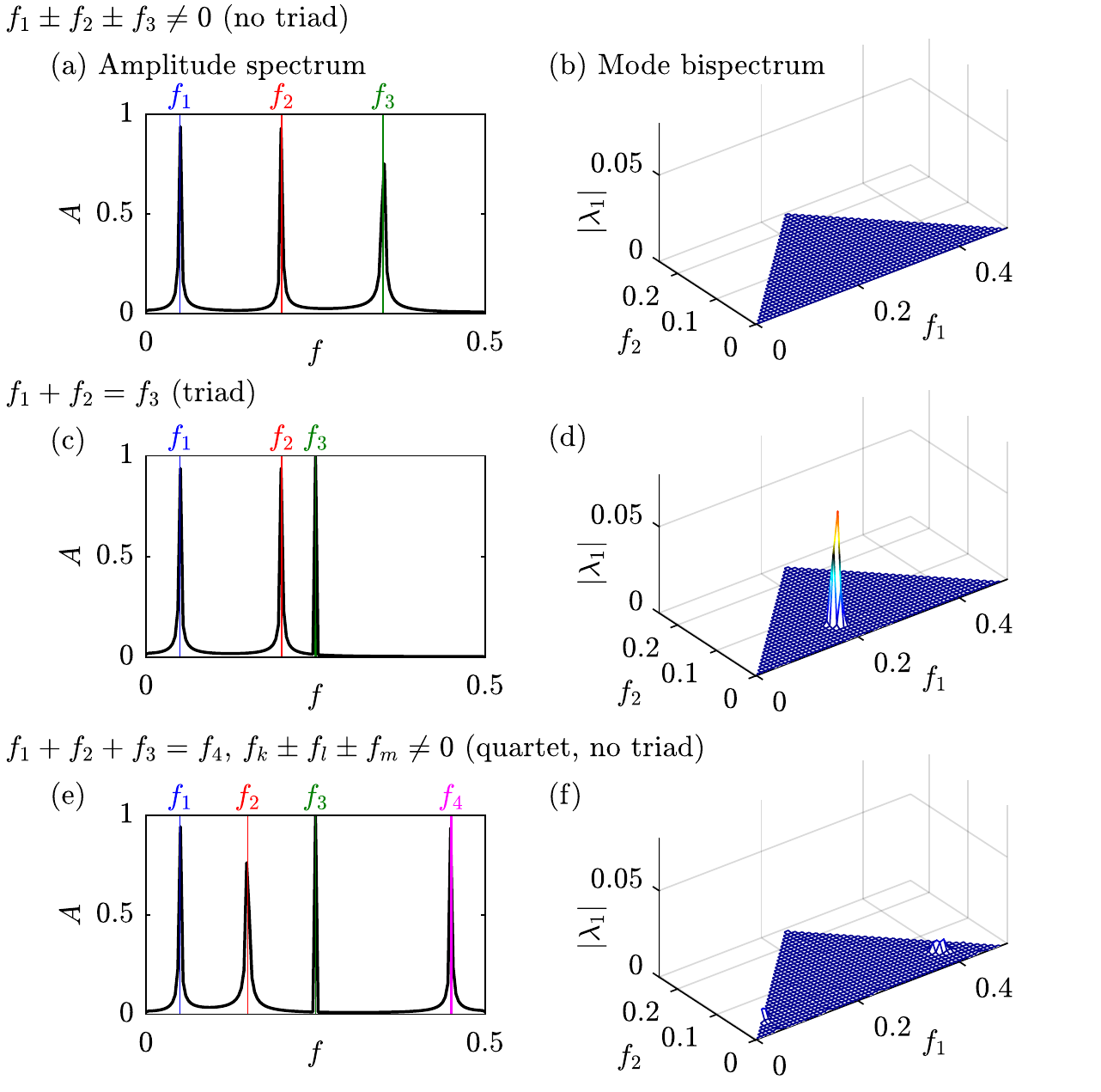}
\caption{Amplitude spectra of ${q}(x=0,t)$ and mode bispectra of the surrogate data: (a,b) non-resonant frequency triplet with $(f_1,f_2,f_3)=(0.05,0.2,0.35)$; (c,d) resonant triad with $(f_1,f_2,f_1 + f_2)=(0.05,0.2,0.25)$; (e,f) resonant four-wave quartet with $(f_1,f_2,f_3,f_1+f_2+f_3)=(0.05,0.15,0.25,0.45)$.}
\label{fig:0hyp_harmonics}       % Give a unique label
\end{figure}
Three test cases are considered in figure \ref{fig:0hyp_harmonics}. For each case, the amplitude spectrum of the time signal at the first point, $A=2|\hat{q}(x=0,f)|$, is compared side-by-side to the mode bispectrum of the full data. The amplitude spectrum is computed using the same spectral estimation parameters as the BMD. The first test shown in figure \ref{fig:0hyp_harmonics}(a,b) demonstrates the rejection of the non-resonant frequency triplet $(f_1,f_2,f_3)=(0.05,0.2,0.35)$ with $f_1+f_2<f_3$. The true negative outcome of the test is apparent from the flat mode bispectrum in \ref{fig:0hyp_harmonics}(b). The presence of spectral leakage can be inferred from the amplitude spectrum, where it affects both a broadening of the spectral peaks and a reduction of their amplitudes. For the second test, $f_3$ is altered such that $f_3=f_1+f_2$. The triplet $(f_1,f_2,f_3)$ now forms a triad and the peak in the mode bispectrum in figure \ref{fig:0hyp_harmonics}(d) clearly indicates its presence. Lastly, we consider the frequency quadruple $(f_1,f_2,f_3,f_4)=(0.05,0.15,0.25,0.45)$ in figure \ref{fig:0hyp_harmonics}(e,f). By letting $f_4=f_1+f_2+f_3$, these four frequencies meet the condition for four-wave resonance, but do not form a triad in any pemutation. As anticipated, the mode bispectrum in figure \ref{fig:0hyp_harmonics}(f) is flat; only small elevations resulting from spectral leakage are observed. The maximum value of $|\lambda_1(f_1,f_2)|$, however, remains an order of magnitude below the peak observed for the true positive test outcome in the presence of a triad interaction in figure \ref{fig:0hyp_harmonics}(d). We hence conclude that the mode bispectrum correctly signals the absence of triads in the data. 

\begin{figure}
  \includegraphics[trim={0cm 0cm 7cm 0cm},clip,width=0.45\textwidth]{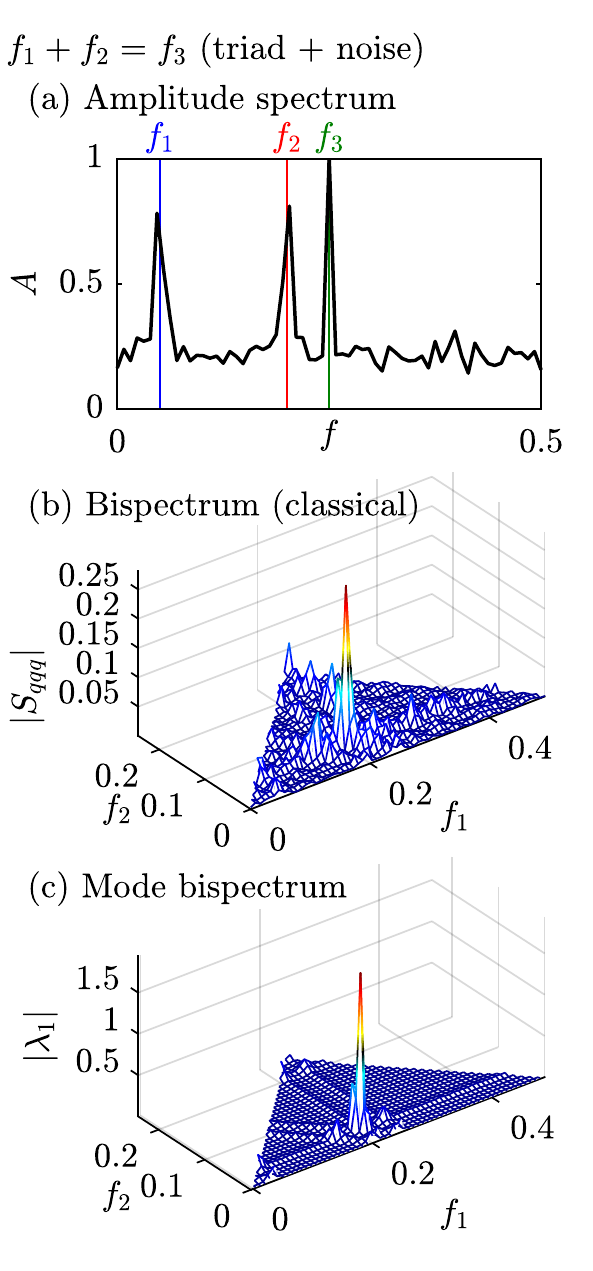}
\caption{BMD of surrogate data in the presence of noise with unit signal-to-noise ratio: (a) amplitude spectrum for ${q}(x=0,t)$; (b) classical bispectrum for ${q}(x=0,t)$; (c) mode bispectrum.}
\label{fig:0hyp_noise}       % Give a unique label
\end{figure}
Next, we test the performance of BMD in the presence of noise. As an example, we revisit the case of a single frequency triad shown in figure \ref{fig:0hyp_harmonics}(c,d) with additive Gaussian white noise. The noise is randomly generated and scaled such that the signal-to-noise ratio is equal to one. This implies that the variance of the noise is the same as the variance of the signal. Signals with a signal-to-noise ratio below this threshold are typically deemed unusable. The presence of noise in the new test data is apparent from the elevation of the amplitude spectrum shown in figure \ref{fig:0hyp_noise}(a). It is at this point instructive to also consider the classical bispectrum defined in equation \eqn{eqn:Sqqq_cont}, which we compute from the same time series as the amplitude spectrum (the classical bispectrum performs the same as the mode bispectrum for the non-noisy data analyzed in figure \ref{fig:0hyp_harmonics}). The susceptibility of the classical bispectrum to noise becomes apparent in figure \ref{fig:0hyp_noise}(b). The bispectrum correctly identifies the triad at $(0.05,0.2)$, but also exhibits a number of lower peaks. These other peaks are attested to the variance of the bispectrum. The mode bispectrum reported in figure \ref{fig:0hyp_noise}(c) clearly indicates the presence of a single triad at the correct frequencies. No significant side peaks are observed. We speculate that this ability to widely reject noise results from the use of spatial correlation information and the optimality property of the decomposition. It is, however, observed that the presence of noise leads to the formation of low-amplitude bands along lines of constant frequencies $f_1$, $f_2$ and $f_3$ (recall that $f_3=f_1+f_2$ is constant along diagonal lines of slope -1). This phenomenon is equally present in figure \ref{fig:0hyp_noise}(b), where it is mostly overshadowed by the larger variance of the classical bispectrum. Similar to spectral leakage in conventional Fourier analysis, these bands are a well-known and commonly ignored phenomenon in bispectral analysis. After establishing that the basic premise of BMD holds for the test data, we proceed by applying BMD to nonlinear flow data in \S \ref{sec:examples}, and refer to appendix \ref{convergence} for an assessment of the convergence of the method.

\section{Examples}\label{sec:examples}
\begin{table}
\centering\label{tab:data}
\begingroup
\setlength{\tabcolsep}{3pt}
\begin{tabular}{lcccccccccc}
Case 		& Variables & $N_x$ & $N_{y,r}$ & $N_{z,\theta}$ & $N_t$ 	& $\Delta t$	& $\Nfft$ 	& $\Novlp$ & $\Nblk$ & $\tol$	\\ \hline
Test data		& $q$	& 100 &	-  & -    & 1280 	& 1		    & 128	    & 0		& 10 & $10^{-8}$ \\
Cylinder DNS		& $u,v$	& 250 &	125  & 1    & 4096 	& 0.06		    & 1024	    & 512		& 7 & $10^{-8}$ \\
Plate PIV		& $u,v$	& 120 &	69  & 1    & $2.5\cdot10^4$ 	& 0.002\it{s}	    & 5000	    & 0		& 50 & $10^{-8}$ \\
Jet LES		& $p$	& 219 &	42 & 128     & $1\cdot10^4$ 	& 0.2		    & 256	    & 128		& 77 & $10^{-15}$ \\
\end{tabular}
\endgroup
\caption{
Parameters of the example databases and spectral estimation parameters. The DNS and LES data are non-dimensionalized by the cylinder diameter and freestream velocity, and the jet diameter and jet velocity, respectively. The PIV data are given in SI units. $\tol$ is the tolerance used by the algorithm presented in appendix \ref{algorithm}. A standard Hann window is used in all cases to reduce spectral leakage.}
\label{tab:data}
\end{table}
In what follows, we conduct BMD analyses of the three representative nonlinear flows summarized in table \ref{tab:data}. The goal of this section is to demonstrate different aspects of BMD using different data---an exhaustive discussion of the nonlinear flow physics of each of these flows is beyond the scope of this work. The dimensionless frequency used in the presentation of the results corresponds to a Stouhal number, but we retain the symbol $f$ for readability. The trade-offs involved in choosing the spectral estimation parameters $\Nfft$ and $\Novlp$ and the windowing function are similar to those for SPOD. The reader is referred to \citet{SchmidtColonius_2020_AIAAJ} for best practices that in large parts translate to BMD. Appendix \ref{convergence} demonstrates the convergence of the results in terms of the summed mode spectra for all three cases.

\subsection{Cylinder flow}\label{sec:cylinder}

\begin{figure}
  \includegraphics[width=0.5\textwidth,trim=0 3.9cm 6.5cm 0cm,clip]{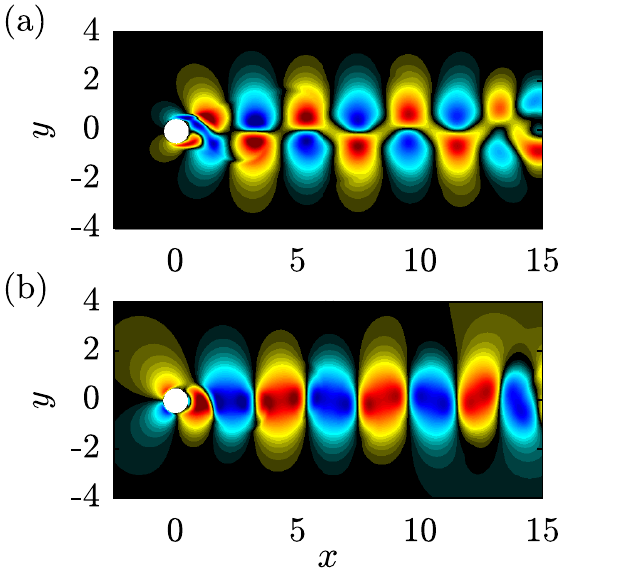}
\caption{Instantaneous fluctuating flow field behind a cylinder at $\Rey=500$: (a) streamwise velocity; (b) transverse velocity. Colormap is saturated at $\pm75\%$ of the free-stream velocity.}
\label{fig:cylinder_inst_uv}       % Give a unique label
\end{figure}
The flow over a cylinder at a Reynolds number, based on the cylinder diameter and the free-stream velocity, of $\Rey=500$ is a canonical laminar, planar flow that exhibits well-understood nonlinear dynamics \citep{williamson1996vortex}. The immersed-boundary solver by \citet{goza2017strongly} was used to solve the incompressible Navier-Stokes equations for the state vector $\vb{q}=[u,v]^T$, consisting of the streamwise and transverse velocity components. Prior to saving the data, the simulation was run for multiple flow-through times to guarantee that the database reported in table \ref{tab:data} represents the limit-cycle solution. The instantaneous flow field is  visualized in figure \ref{fig:cylinder_inst_uv}.

\begin{figure}
  \includegraphics[width=\textwidth]{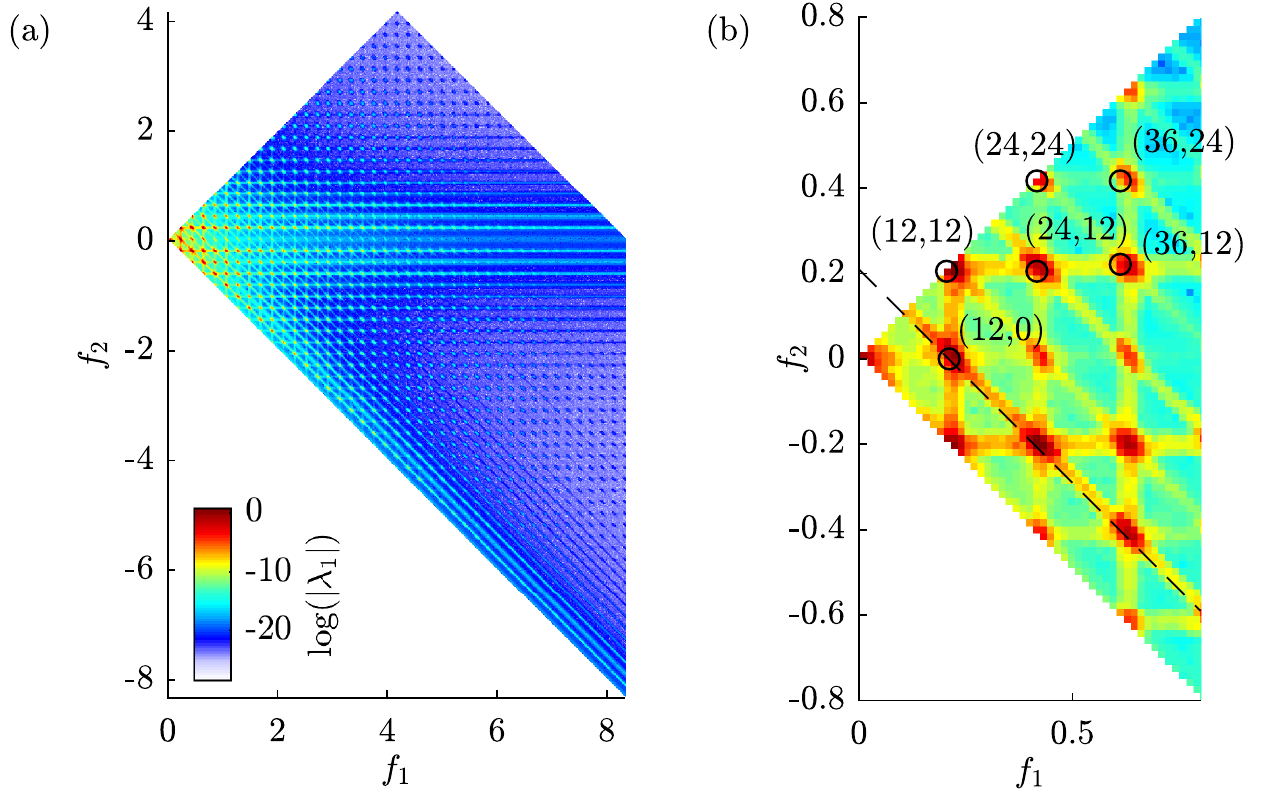}
\caption{(Magnitude) mode bispectrum for cylinder flow at $\Rey=500$: (a) mode bispectrum for $\Nfft=1024$ in the sum and difference regions; (b) magnification of the low frequency region. A cascade of triads ($\circ$) that originates from the fundamental instability and the constant frequency $f_0=0.21$ (- - -) of the fundamental instability are marked in (b).}
\label{fig:cylinder_bispectrum}       % Give a unique label
\end{figure}
Figure \ref{fig:cylinder_bispectrum}(a) shows the mode bispectrum and figure \ref{fig:cylinder_bispectrum}(b) a magnification of the low-frequency portion. The most striking feature is a distinct grid pattern with local maxima at its nodes. These local maxima are the footprint of a cascade of triads that is generated through the mechanism illustrated in figure \ref{fig:principal}(b). As briefly discussed in \S \ref{sec:test}, the horizontal, vertical and diagonal bands observed in the mode bispectrum results from spectral leakage. This phenomenon is inherent to the discrete Fourier transform of non-periodic data. It is not specific to BMD and should not be physically interpreted. A closer inspection of figure \ref{fig:cylinder_bispectrum}(b) reveals that the global maximum of the mode bispectrum occurs for the index doublet $(k,l)=(12,12)$, that is, the triad $(k,l,k+l)=(12,12,24)$ (or $\{f_1,f_2,f_1+f_2\}=\{0.21,0.21,0.42\}$ in terms of frequency). This maximum corresponds to the sum-interaction of the fundamental instability with itself, $\{f_0,f_0,2f_0\}$, which generates the first harmonic at twice that frequency, that is, $2f_0=0.42$. The difference-interaction of the fundamental instability with itself, $\{f_0,-f_0,0\}$, on the other hand, leads to a mean flow deformation that is indicated by the maximum at $(12,0)$ on the $f_1$-axis. Sum- and difference-self-interactions are illustrated in figure \ref{fig:triadic_interaction_Feynman}(d) and \ref{fig:triadic_interaction_Feynman}(c), respectively. The detection of the fundamental difference-self-interaction, $\{f_0,-f_0,0\}$, has an important implication for the interpretation of the mode bispectrum: in addition to triad interactions, it indirectly detects intrinsic instability mechanisms that, after attaining an appreciable amplitude, nonlinearly self-interact to generate a mean flow distortion. Here, this instability mechanism is the bluff-body vortex shedding behind the cylinder.

\begin{figure}
  \includegraphics[width=\textwidth,trim=0 1.6cm 0cm 0.8cm,clip]{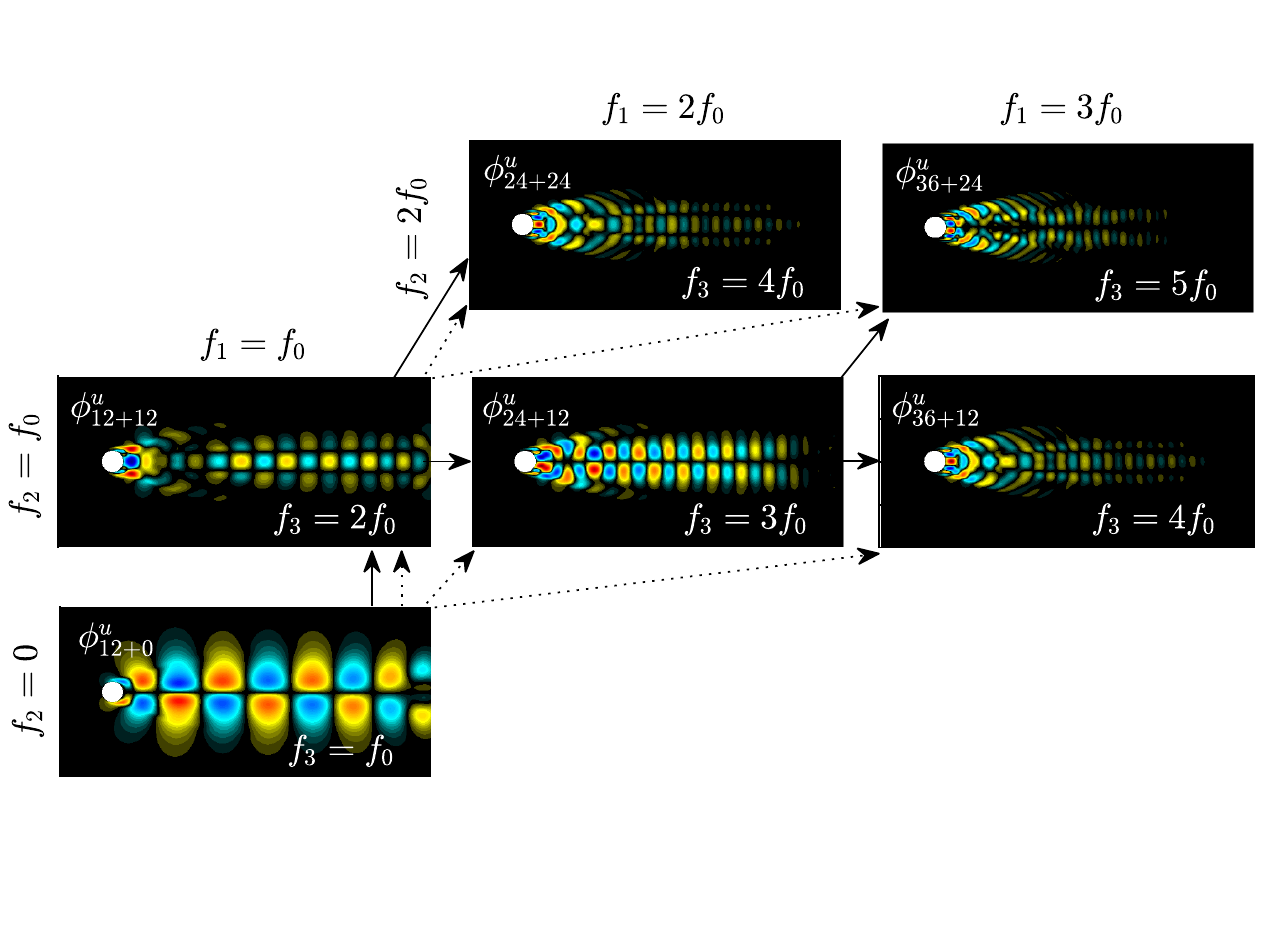}
\caption{Bispectral modes forming a cascade of triads, marked in figure \ref{fig:cylinder_bispectrum}, starting from the fundamental instability at $f_0$ (bottom-left). Arrows indicate the triadic interactions ($\rightarrow$: $k$-component, $\dashrightarrow$: $l$-component). The real part of the streamwise velocity component is shown.}
\label{fig:cylinder_umodes_grid}       % Give a unique label
\end{figure}
The spatial structures of the bispectral modes associated with the cascade of triads previously marked by circles in figure \ref{fig:cylinder_bispectrum}(b), are shown in figure \ref{fig:cylinder_umodes_grid}. The solid and dotted arrows indicate the sum-interactions of the $k$-th and $l$-th frequency components, respectively, that generate the $k+l$-th component. The cascade starts with the fundamental mode $\phi_{12+0}$. Its self-interaction generates mode $\phi_{12+12}$, which in turn partakes in the generation of modes $\phi_{24+12}$ and $\phi_{24+24}$, and so on. The spatial structures of the modes reveals that each interaction yields new wavenumber components in the streamwise and/or transverse directions. 

\begin{figure}
  \includegraphics[width=\textwidth,trim=0 3cm 0cm 1.25cm,clip]{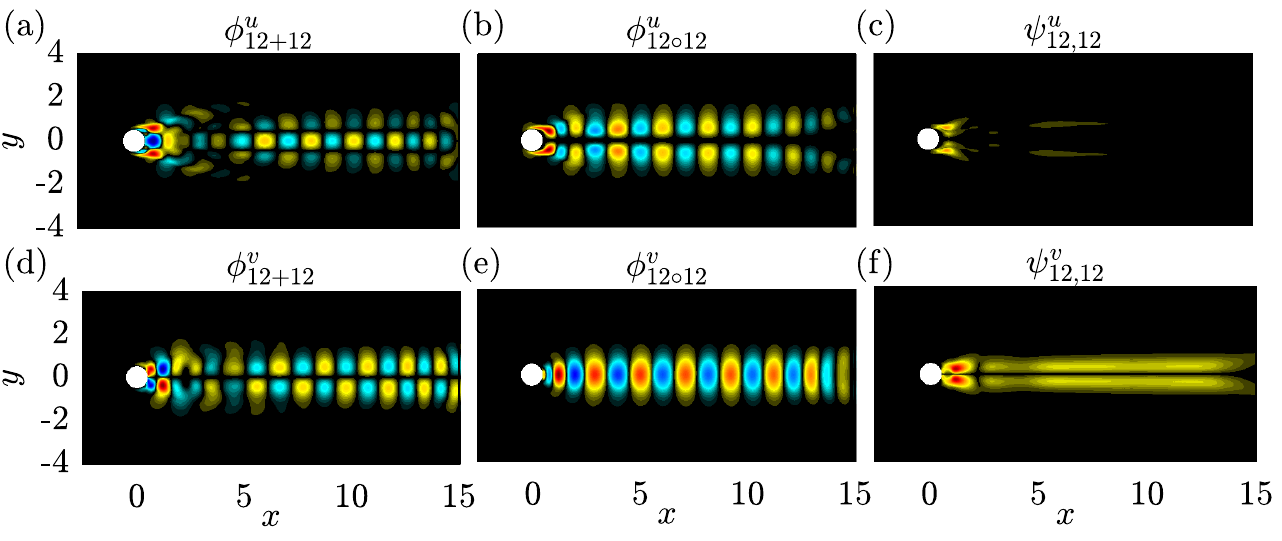}
\caption{Self-interaction of the fundamental mode: (a,d) bispectral mode; (b,e) cross-frequency field; (c,f) interaction map. Top and bottom rows show the $u$ and $v$ components, respectively. False colors of modes are saturated at maximum absolute value. False colors of interaction maps are identical to allow for comparison. The real part of the bispectral modes is shown in (a,b,d,e).}
\label{fig:cylinder_interactionmap}   % Give a unique label
\end{figure}
The self-interaction of the fundamental is investigated in more detail in figure \ref{fig:cylinder_interactionmap}. The $u$- and $v$-components of the corresponding bispectral mode and cross-frequency field are reported in the upper and lower row, respectively. Following the definition in equation \eqn{eqn:interaction_map}, the entry-wise product of the bispectral mode and the cross-frequency field yields the interaction map shown in figure \ref{fig:cylinder_interactionmap}(b,e). The main observations is that the interaction is the strongest in the wake region just downstream of the cylinder. The transverse component furthermore attains a larger maximum value than the streamwise component and is less spatially confined. A connection to the sensitivity regions identified experimentally by \citet{strykowski1990formation}, and predicted based on structural stability analysis by \citet{giannetti2007structural}, remains speculative.

%\newpage
%\clearpage
\subsection{Massively-separated flow behind flat plate at high angle of attack}\label{sec:plate}
The second example is that of PIV data of massively-separated flow behind flat plate at high angle of attack. A total of 50 independent measurements consisting of 5000 snapshots each is used. Unlike the cylinder flow simulation data, these data are subject to measurement noise and stochasticity, and exhibit much richer dynamics.
\begin{figure}
  \includegraphics[width=\textwidth,trim=0 4.5cm 0cm 0cm,clip]{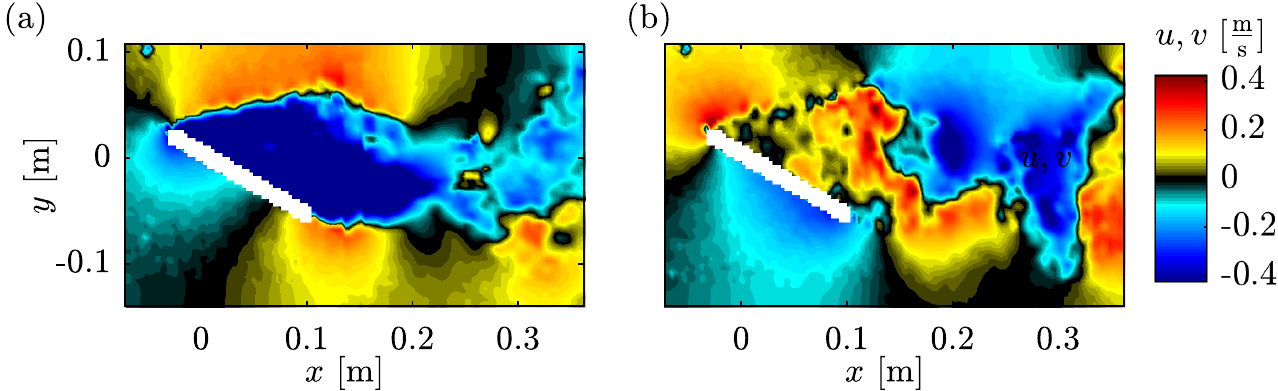}
\caption{Instantaneous PIV flow field of massively-separated flow behind a flat plate: (a) streamwise fluctuating velocity; (b) transverse fluctuating velocity. (Data courtesy of K. Mulleners, EPFL.)}
\label{fig:plate_inst_uv}       % Give a unique label
\end{figure}
This becomes apparent from the instantaneous flow field visualization in figure \ref{fig:plate_inst_uv}. 

\begin{figure}
  \includegraphics[width=\textwidth,trim=0 0.1cm 0cm 0cm,clip]{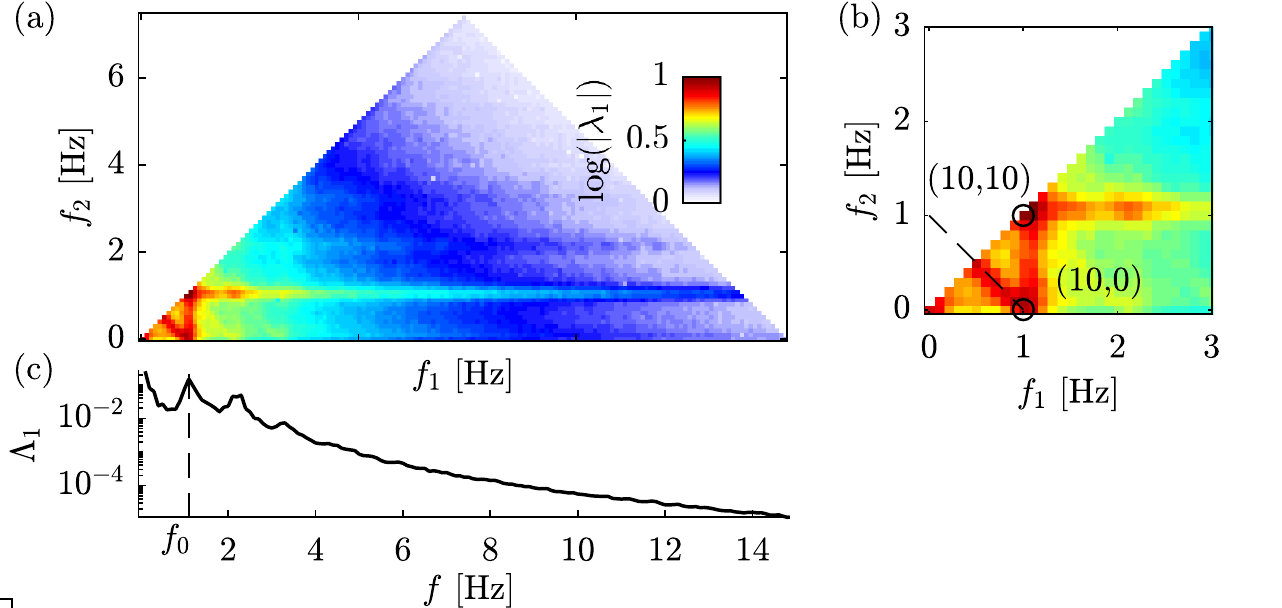}
\caption{Spectra of PIV data of massively-separated flow behind plate: (a) mode bispectrum for $\Nfft=5000$; (b) magnification of the bottom-left corner; (c) summed mode spectrum. The fundamental vortex-shedding frequency is $f_0=1.0\:$Hz (frequency index 10) is marked by dashed lines (- - -) in (a,b,c). The global maximum of the mode bispectrum occurs for the self-interaction triad $(10,10)$ of the fundamental instability.}
\label{fig:plate_spectra}       % Give a unique label
\end{figure}
Focusing again on the sum-interaction region, the magnitude and summed mode bispectra are shown in figure \ref{fig:plate_spectra}. Similar to figures \ref{fig:cylinder_bispectrum} and \ref{fig:cylinder_sumspectrum} for the cylinder flow, the mode and summed mode bispectra unveil the signature of bluff-body vortex shedding with the self-interaction triad of the fundamental instability at its center.

\begin{figure}
  \includegraphics[width=\textwidth,trim=0 8cm 0cm 0cm,clip]{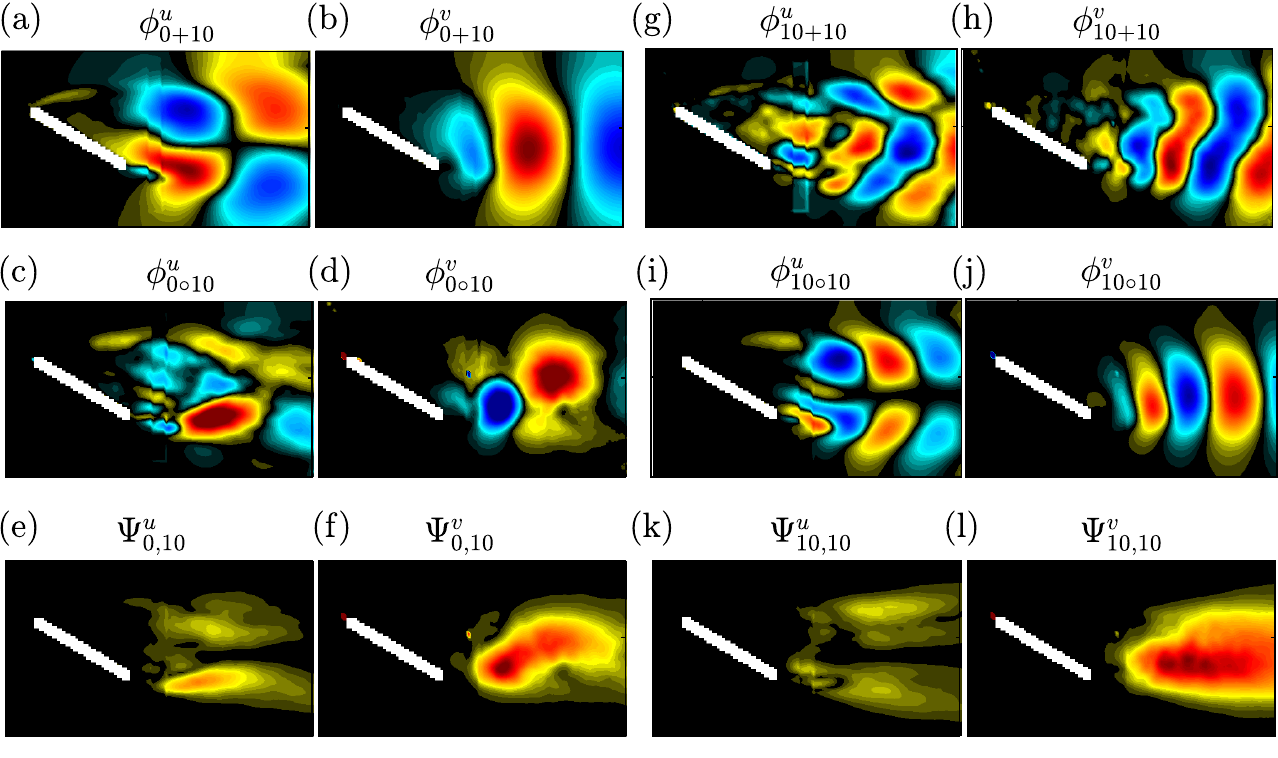}
\caption{Streamwise and transverse components of modes of PIV data of massively-separated flow behind plate: (a-f) fundamental instability; (g-l) self-interaction of fundamental instability. Top, middle and bottom rows show bispectral modes, cross-frequency fields, and interactions maps, respectively. The real part of the bispectral modes is shown in (a-d,g-j). The vertical bar particularly visible in (c,g) is an artifact of two overlapping PIV windows.}
\label{fig:plate_modes}       % Give a unique label
\end{figure}
As before, we analyze the fundamental self-interaction triad in more detail by inspecting the corresponding bispectral modes, cross-frequency fields and interaction maps. This is done in figure \ref{fig:plate_modes}. The fundamental vortex shedding mode shown in figure \ref{fig:plate_modes}(a,b) becomes symmetric at a short distance downstream from the plate. The mode generated by the self-interaction of the fundamental, mode $(10,10)$, is shown in figure \ref{fig:plate_modes}(g,h). From inspecting and comparing the interaction maps shown in figure \ref{fig:plate_modes}(k,l), we conclude that the triadic interaction takes place predominantly in the transverse velocity component and in the bottom shear layer. For both the cylinder flow and the massively-separated plate, we confirmed that the fundamental modes, (12,0) and (10,0), respectively, correspond to the overall most energetic large-scale coherent structures as identified by SPOD. In fact, the fundamental bispectral modes shown in figures \ref{fig:cylinder_umodes_grid} (bottom-left) and \ref{fig:plate_modes}(a,b) are almost indistinguishable from the most energetic SPOD modes (not reported here for brevity). The structures of the leading modes are clearly identified as symmetric and anti-symmetric wave trains, despite the stochastic and nature of the data. Comparing the modes with those obtained for laminar cylinder flow suggests that the laminar bluff-body dynamics prevail in the turbulent regime. It is understood that symmetries that are broken at low Reynolds numbers, and flow structures that resemble laminar instability modes in the same regime, resurface in fully developed turbulent flows at very high Reynolds numbers. Quantitative empirical evidence for this phenomenon in the wake of an axisymmetric bluff body, for example, was provided by \citet{rigas2014low}.

%\newpage
%\clearpage
\subsection{Jet at $\Rey=3600$} \label{sec:jet}
The example of an initially laminar jet at a moderate Reynolds number of $\Rey=3600$ is chosen to demonstrate the treatment of flows with homogeneous directions, here the azimuthal direction in a cylindrically symmetric domain. The large eddy simulation was conducted by Dr.~G.~A. Br\`es using the numerical framework discussed in \citet{bres2019modelling}. The original data was computed on an unstructured grid. The database used here was later interpolated onto a cylindrical grid with coordinates $\vb{x}=[x,r,\theta]^T$, where $r$ and $\theta$ are the radial and azimuthal coordinates, respectively. We may exploit the cylindrical symmetry of the jet by decomposing the flow field into azimuthal Fourier modes
\begin{equation}
        \tilde{\vb{q}}(x,r,m,t) = \sum_{j=0}^{N_\theta-1}\vb{q}(x,r,\theta_{j+1},t)\ee^{-\ii m}, \quad  m=0,\dots,N_\theta-1 \label{eqn:fft_m}
\end{equation}
of azimuthal wavenumber $m$. Following the discussion in \S\ref{sec:space_hom}, we consider azimuthal triads which we denote by triplets $\ldb m_1,m_2,m_3\rdb$, where $m_3=m_1+m_2$. We denote by $\ldb\cdot\rdb$ azimuthal wavenumber multiplets, to avoid confusion. Since the jet has no preferred sense of rotation, the azimuthal wavenumber spectrum is expected to be symmetric and it suffices to consider positive azimuthal wavenumbers $m\geq0$. In the light of rotational symmetry, equation \eqn{eqn:homdir} specializes as
\begin{equation}\label{eqn:homdir_jet}
\vb{B}\left(\mqty[x\\r\\ \theta],\mqty[x'\\r'\\ \theta'],f_k,f_l\right) \rightarrow \vb{B}\left(\mqty[x\\ r],\mqty[x'\\ r'],m_i,m_j,f_k,f_l\right).
\end{equation}
In the following, we will conduct a bispectral mode analysis on the fluctuating pressure field and restrict our attention, for brevity, to azimuthal wavenumber combinations with $m_3\leq3$.

\begin{figure}
  \includegraphics[width=\textwidth,trim=0 5cm 0cm 0cm,clip]{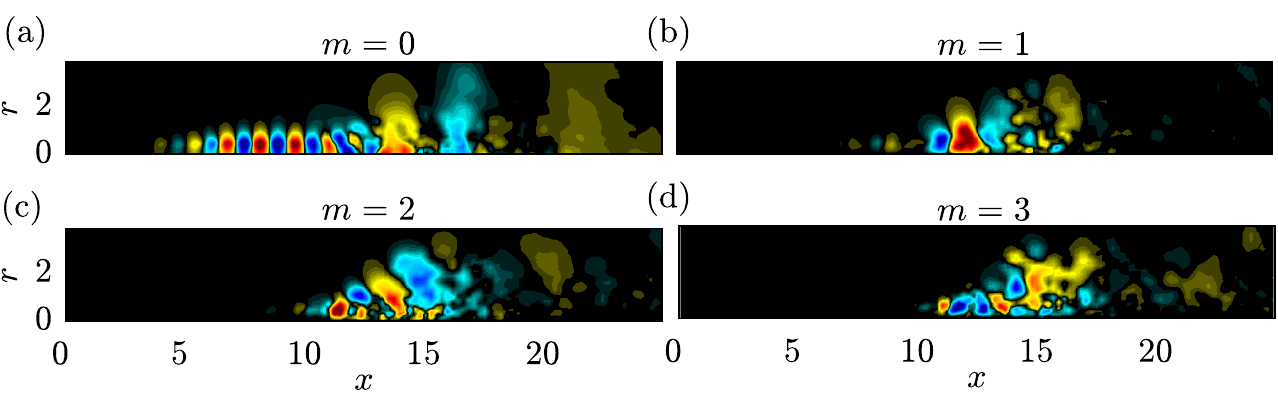}
\caption{Instantaneous pressure fields of the leading azimuthal wavenumber components of the transitional round jet: (a) $m=0$; (b) $m=1$; (c) $m=2$; (d) $m=3$. False colors are saturated at maximum absolute value of each component. For $m>0$, the real part of the pressure field is shown.}
\label{fig:jet_inst_p}       % Give a unique label
\end{figure}
 The instantaneous fluctuating pressure fields of the first four azimuthal wavenumber components are shown in figure \ref{fig:jet_inst_p}. From figure \ref{fig:jet_inst_p}(a), it can be seen that the annular shear-layer supports a symmetric Kelvin-Helmholtz instability that breaks down into turbulence at $x\approx12$. Comparison with figure \ref{fig:jet_inst_p}(b-d) suggests that this breakdown is three-dimensional and leads energy transfer to higher azimuthal wavenumber components. It is this interaction across azimuthal wavenumbers that we study using BMD in the following. 

\begin{figure}
  \includegraphics[width=\textwidth,trim=0 3.75cm 0cm 0cm,clip]{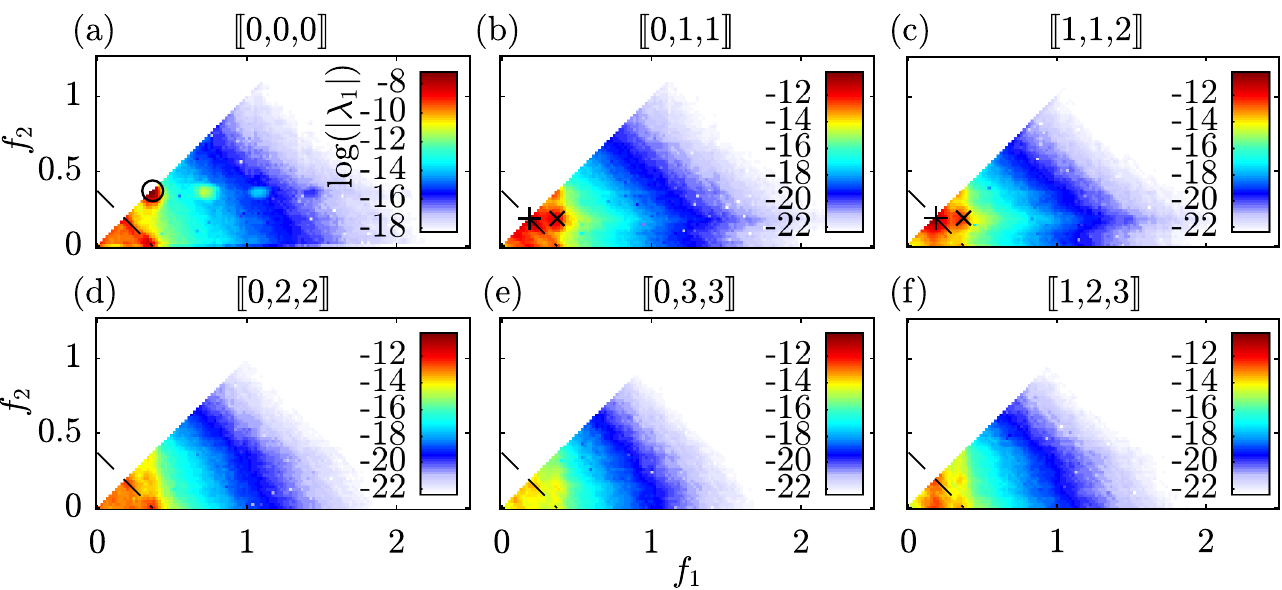}
\caption{sum-interaction regions of mode bispectra for the interaction of different azimuthal wavenumber components of the transitional jet: (a) self-interaction of $m=0$; (b) interaction of $m=1$ and 2; (c) self-interaction of $m=1$; (d) interaction of $m=0$ and 2; (e) interaction of $m=0$ and 3; (f) interaction of $m=1$ and 2. Azimuthal triads are denoted by triplets $\ldb m_1,m_2,m_3 \rdb$. Dashed lines (- - -) mark the diagonal of constant frequency $f_0$. False color ranges in (b-f) are identical and significantly lower than for $\ldb 0,0,0\rdb$. The fundamental self-interaction frequency triad $\{f_0,f_0,2f_0\}$ is marked by `$\circ$' in (a), and the triads $\{\frac{1}{2}f_0,\frac{1}{2}f_0,f_0\}$ and $\{f_0,\frac{1}{2}f_0,\frac{3}{2}f_0\}$ as `$+$' and `$\times$', respectively, in (b,c).}
\label{fig:jet_bispectrum_all}       % Give a unique label
\end{figure}
We will not investigate the modes in detail, but instead focus on the sum-interaction region of the mode bispectra for the six principle triplets with $m_3\leq3$ shown in figure \ref{fig:jet_bispectrum_all}. Maxima in these mode bispectra evidence the presence of azimuthal triads. Given that the jet dynamics are in large parts dominated by the symmetric Kelvin-Helmholtz shear layer roll-up, it comes to no surprise that the fundamental self-interaction occurring for $\ldb 0,0,0\rdb$, $\{f_0,f_0,2f_0\}$, marked by ($\circ$) in figure \ref{fig:jet_bispectrum_all}(a), is associated with the global maximum of the bispectral density. The local maxima observed in \ref{fig:jet_bispectrum_all}(b-f) suggest that azimuthal wavenumber triads at low frequencies cascade energy to higher and higher azimuthal wavenumbers. For the triplets $\ldb0,0,1\rdb$ and $\ldb1,1,2\rdb$ shown in figure \ref{fig:jet_bispectrum_all}(b) and \ref{fig:jet_bispectrum_all}(c), respectively, this interaction takes place at frequencies $\{\frac{1}{2}f_0,\frac{1}{2}f_0,f_0\}$ and $\{f_0,\frac{1}{2}f_0,\frac{3}{2}f_0\}$, i.e., it involves subharmonic frequencies like $\frac{1}{2}f_0$ and leads to the generation of the ultraharmonic frequency components like $\frac{3}{2}f_0$.

\begin{figure}
  \includegraphics[width=\textwidth,trim=0 0cm 0cm 0cm,clip]{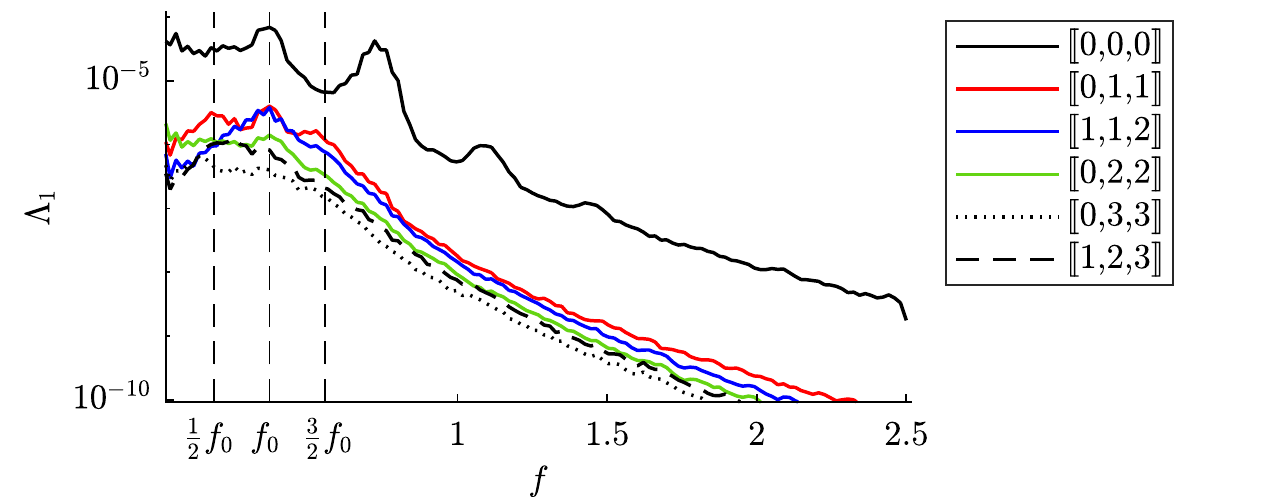}
\caption{Summed mode spectra for the interaction of different azimuthal wavenumber components of the transitional jet. Triplets $\ldb m_1,m_2,m_3\rdb$ denote azimuthal triads. As in figure \ref{fig:jet_bispectrum_all} above, the sum-interaction region is considered.}
\label{fig:jet_sumspec_all}       % Give a unique label
\end{figure}
The summed mode spectra for the same azimuthal wavenumber triplets are compared in figure \ref{fig:jet_sumspec_all}. The summed mode spectrum for the $(0,0,0)$ interaction has a significantly higher base-level that the other spectra and shows peaks at the fundamental frequency and three higher harmonics. The most prominent peak is observed for the first harmonic at $f=2f_0$. For $\ldb 0,1,1\rdb$, peaks on right and left side of the fundamental frequency indicate the presence of the ultraharmonic and the subharmonic frequency components, $\frac{3}{2}f_0$ and $\frac{1}{2}f_0$, respectively. Using BMD, we were able to show that these fractional frequencies are generated by triad interactions between different azimuthal wavenumber components.

\section{Discussion}\label{sec:discussion}

Bispectral mode decomposition was introduced as a means of educing coherent structures associated with triad wave interactions from flow data. The method is based on the maximization problem \eqn{eqn:a1} for the spatio-temporal bispectral density. It was shown that this problem is directly related to the numerical radius of the bispectral density matrix, defined in equation \eqn{eqn:bsd}. Unlike the classical bispectrum, the devised multidimensional decomposition establishes a causal relationship between the three frequency components that comprise a triad. As a result, the method is capable of educing coherent structures involved in sum- and difference-interactions.

Possible extensions of the framework include the analysis of four-wave resonances, a common phenomenon in water waves \citep{hammack1993resonant,janssen2003nonlinear}, based on the trispectrum. Such a trispectral mode decomposition could be based on either the fourth-order moment or the fourth-order cumulant. We speculate that the cumulant is preferable as it excludes contributions from lower-order moments. This distinction was not necessary in the present work since the third-order moment and cumulant are identical. 

Potential future applications include the estimation of nonlinear transfer functions based on the mode (cross-) bispectrum. Starting from the Navier-Stokes equations in spectral form, \cite{domaradzki1990local} computed the nonlinear energy transfer term in the energy amplitude equation directly from data. A statistical method that does not require knowledge of the governing equations, but instead uses bispectral information to identify the linear and quadratic transfer functions of a single-input and single-output system, was proposed by \citet{ritz1986estimation}. In a similar manner, \cite{kim1988digital} informed from data the transfer functions in the second-order Volterra series of a nonlinear time-invariant system. BMD can potentially be leveraged to extend such nonlinear system identification approaches from single-input and single-output to entire flow fields. Another direction of future research are reduced-order models based on modal expansions that only include the dynamically most relevant modes, as identified by the mode bispectrum, and their quadratic interactions. In the context of flow control, the interaction maps defined by equation \eqn{sec:intermaps} can potentially be used to localize, and selectively mitigate, certain triad interactions. A control strategy for mitigating extreme dissipation events triggered by triad interactions in turbulent flows, for example, was recently proposed by \cite{farazmand2019closed}. Combined with BMD-based system identification, as proposed above, this goal can potentially be achieved without knowledge of the governing equations. The effect of actuation on the flow field can further be studied by means of a cross-bispectral mode decomposition between actuation and the flow field, see below.

\section{Conclusions}\label{sec:conclusions}

\sloppypar The decomposition was applied to three experimental and numerical flow databases that represent the laminar, transitional and turbulent regimes. Quadratically interacting frequency components were identified as maxima in the mode bispectrum and the corresponding bispectral modes reveal the flow structures that are generated through the interactions. Two additional quantities were used to aid the analysis and physical interpretation: interaction maps that identify regions of activity of triadic interaction, and the summed mode bispectrum as a compact representation of the mode bispectrum. 

For laminar cylinder flow at $\Rey=500$, cascading triads and their regions of interaction were educed and it was demonstrated that difference-self-interactions that entail distortions of the mean flow indirectly reveal the intrinsic vortex-shedding mechanism of this flow. Applicability to turbulent flows and in the presence of measurement noise were demonstrated on particle image velocimetry data of massively-separated flow behind a flat plate at high angle of attack. In large eddy simulation data of a transitional round jet at $\Rey=3600$, the generation of sub- and ultraharmonics was explained by extending the method to incorporate cross-bispectral information between different azimuthal wavenumber components.

\paragraph{\bf Acknowledgements.} I would like to thank Tim Colonius for pointing me to higher-order spectra, Ethan Pickering for pointing out the need to compute difference-interactions, Aaron Towne, Peter Schmid, Georgios Rigas and Tim Colonius for many helpful discussions, and the two anonymous referees for their very insightful comments. I gratefully acknowledge Karen Mulleners for providing the PIV data, Guillaume Br\`es for the LES data, and Andres Goza for helping with the DNS.

%\paragraph{\bf Declaration of Interests.} The author reports no conflict of interest.
\newpage
\clearpage

%\begin{acknowledgements}
%\end{acknowledgements}
%\newpage
\appendix
\section{Computation of $\vb{a}_1= \argmax_{\|\vb{a}\|=1}\qty|\vb{a}^*\vb{B}\vb{a}|$  } \label{algorithm}
A slightly modified version of the algorithm for the computation of the numerical radius by \cite{he1997algorithm} is used. He and Watson's algorithm requires two nested iterations. The first, or so-called simple iteration \citep{watson1996computing}, converges to a local solution of equation (\ref{eqn:maxtheta}). A tolerance of $\tol=10^{-8}$ was found to be a good compromise between accuracy and compute time for both iterations. The number of iterations was limited to $k=300$. 
	
\vspace{2mm}{\bf\noindent Algorithm 1} (Watson's Simple Iteration) \\
\noindent\begin{algorithm}[H]
\SetAlgoLined
\KwData{$\vb{B}$, $\tol$ and $\vb{a}_0$, a complex vector as initial guess.}
\KwResult{$\vb{a}$ and $w$, a local solution of equation (\ref{eqn:maxtheta}).}
\For{$k=1,2,\dots$}{
	Set $w_{k-1}=\vb{a}_{k-1}^H\vb{B}\vb{a}_{k-1}$. \\
	Define $\vb{a}_k$  by 
	\[\vb{a}_k=w_{k-1}\vb{B}^H\vb{a}_{k-1} + w_{k-1}^H\vb{B}\vb{a}_{k-1}.\] \\
	Normalize $\vb{a}_k\leftarrow\frac{\vb{a}_k}{\|\vb{a}_k\|}$. \\
	Return $\vb{a}=\vb{a}_k$ and $w=\vb{a}_{k}^H\vb{B}\vb{a}_{k}$ if $|w_{k-1}-w_{k-2}|<\tol$.	
	}
% \caption{How to write algorithms}
\end{algorithm}

Building on the simple iteration to find local solutions, the purpose of the main algorithm is to find the global solution. Double precision arithmetic with machine precision $\epsilon=2^{-52}$ was used to compute the results in \S \ref{sec:examples}. If the algorithm did not converge within 500 iterations, it was restarted up to five times with a new random initial guess for $\vb{a}_0$. This procedure was necessary to ensure that all results are fully converged.

\vspace{2mm}{\bf\noindent Algorithm 2} (He and Watson's Algorithm)	\\
\begin{algorithm}[H]
\SetAlgoLined
\KwData{$\vb{B}$ and $\tol$.}
\KwResult{$\vb{a}\approx\vb{a}_1$, $\lb\approx r(\vb{B})$, $\ub$. $\lb$ and $\ub$ are a lower bound and an upper bound of $r(\vb{B})$, such that $\ub-\lb\leq\tol$.}

Set $\vb{a}_0$ to a random complex vector. \\
Set $\lb = 0$ and $\ub = \|\vb{B}\|_1$, the matrix 1-norm of $\vb{B}$. \\
\While{$\ub - \lb > \tol$}{
Use Algorithm 1 with starting vector $\vb{a}_0$ to obtain an updated vector $\vb{a}$. \\
Set $\lb = \max(\lb, |\vb{a}^H\vb{B}\vb{a}|)$.\\
Set $\alpha = \lb + \tol$ and solve generalized eigenvalue problem 
\[
\vb{R}(\alpha)\vb{v} = \lambda\vb{S}\vb{v},
\]
where
\[
\vb{R}(\alpha) = \mqty[2\alpha\vb{I} & -\vb{B}^H \\ \vb{I} & \vb{0}], \quad \vb{S} = \mqty[\vb{B} & \vb{0} \\ \vb{0} & \vb{I}],
\] and $\vb{I}$ is the $\Nblk\times\Nblk$ identity matrix.\\
\eIf{$|\lambda-1|<\sqrt{\epsilon}\|\vb{B}\|_1 \;\; \mathrm{ for\;any }\; \lambda$}{(i.e., there is no eigenvalue on the unit circle)\\
Set $\ub = \lb + \tol$ and return $\vb{a}$, $\lb$, and $\ub$. \\
}{
Set $\vb{a}$ equal to the last $\Nblk$ components of an eigenvector $\vb{v}$ corresponding to an eigenvalue on the unit circle.\\
}
}
%\caption{How to write algorithms}
\end{algorithm}	
	
\section{Convergence}\label{convergence}
\begin{figure}
  \includegraphics[width=0.8\textwidth]{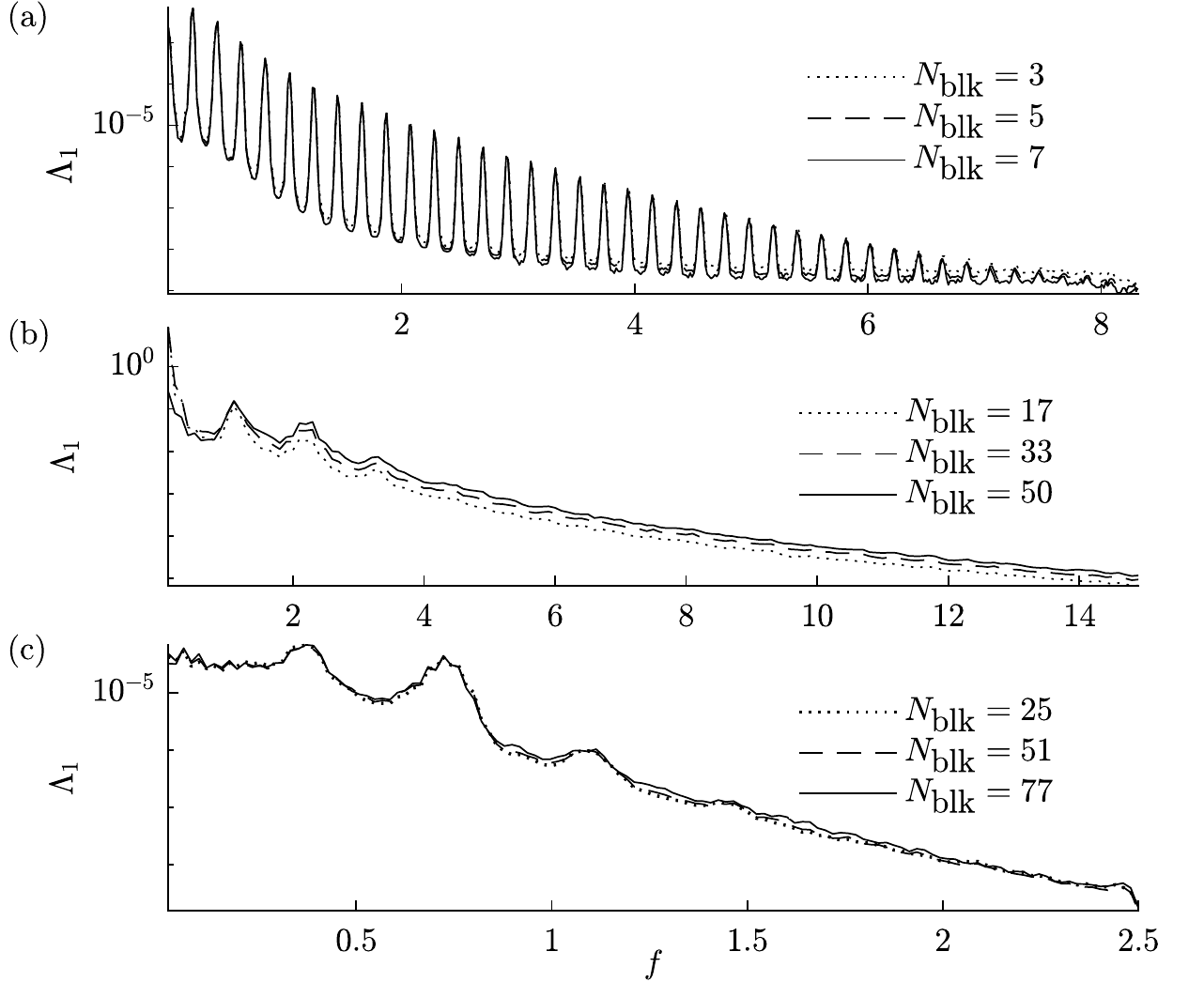}
\caption{Convergence of summed mode spectra for different $\Nblk$: (a) cylinder DNS; (b) flat plate PIV; (c) $\ldb 0,0,0\rdb$ azimuthal triplet of the transitional jet LES. Solid lines correspond to the full data.}
\label{fig:sumspec_convergence}       % Give a unique label
\end{figure}
Figure \ref{fig:sumspec_convergence} demonstrates the convergence of all three cases discussed in \S \ref{sec:examples} in terms of summed mode spectra. The convergence of the results is tested by recomputing the BMD for smaller subsets of the available data. In particular, figure \ref{fig:sumspec_convergence}a shows that the summed mode spectrum obtained for the first 3 and 5 blocks of the cylinder flow DNS data is very similar to the one obtained for all 7 block. Similarly, the full data of the flat plate PIV and the jet LES are compared to the spectra obtained for approximately one-third and two-thirds of the full data in \ref{fig:sumspec_convergence}b and \ref{fig:sumspec_convergence}c, respectively. It can be seen that the summed mode spectra for the latter cases, too, remain largely unaltered by the data reduction. The summed mode spectrum was chosen as the most compact representation of the results. A comparison of the bispectral modes and mode bispectra (not shown) obtained from the full and reduced data sets confirmed that the results are also well-converged with respect to mode spectra and mode shapes.

%The summed mode bispectrum  $\Lambda_1(f)$, i.e., the average bispectral density of all frequencies $f_1$ and $f_2$ that sum to any given frequency $f$, is presented in figure \ref{fig:cylinder_sumspectrum}. A number of peaks clearly identify the fundamental shedding frequency and its higher harmonics. In accordance with the aforementioned dominance of the fundamental instability, the summed mode bispectrum also attains its maximum at the corresponding frequency, $f_0$.

\section*{Conflict of Interest}
The authors declare that they have no conflict of interest.
	
\bibliographystyle{spbasic}      % basic style, author-year citations
%\bibliographystyle{abbrvnat}
\bibliography{/Users/oschmidt/Documents/Paper/DG,/Users/oschmidt/Documents/Paper/PODetc,/Users/oschmidt/Documents/Paper/PSE_and_LST,/Users/oschmidt/Documents/Paper/acoustics,/Users/oschmidt/Documents/Paper/books,/Users/oschmidt/Documents/Paper/cavitation,/Users/oschmidt/Documents/Paper/climate,/Users/oschmidt/Documents/Paper/corner_flow,/Users/oschmidt/Documents/Paper/jets,/Users/oschmidt/Documents/Paper/machine-learning,/Users/oschmidt/Documents/Paper/machineLearning,/Users/oschmidt/Documents/Paper/non-modalStab,/Users/oschmidt/Documents/Paper/others,/Users/oschmidt/Documents/Paper/stochastics,/Users/oschmidt/Documents/Paper/turbulence_and_receptivity,/Users/oschmidt/Documents/Paper/hypersonics}
%\begin{thebibliography}{}

%\end{thebibliography}

\end{document}